\documentclass[11pt]{article}
\pdfoutput=1

\usepackage{fullpage}
\usepackage{amsmath}
\usepackage{amssymb}
\usepackage{setspace}
\usepackage{bbm}
\usepackage{dsfont}
\usepackage{graphics}

\usepackage{color}

\usepackage[colorlinks=true]{hyperref} 
\hypersetup{
    bookmarks=true,         
    unicode=false,          
    pdftoolbar=true,        
    pdfmenubar=true,        
    pdffitwindow=false,     
    pdfstartview={FitH},    
    pdftitle={My title},    
    pdfauthor={Author},     
    pdfsubject={Subject},   
    pdfcreator={Creator},   
    pdfproducer={Producer}, 
    pdfkeywords={keyword1} {key2} {key3}, 
    pdfnewwindow=true,      
    colorlinks=true,       
    linkcolor=blue,          
    citecolor=red,        
    filecolor=magenta,      
    urlcolor=cyan           
}

\usepackage[normalem]{ulem}
\usepackage{amsmath}
\usepackage{enumerate}
\usepackage{amsfonts}
\usepackage{yfonts}

\usepackage{subfigure}
\usepackage{psfrag}

\usepackage{epsfig}
\usepackage[latin1]{inputenc}
\usepackage{float}
\usepackage{graphicx}
\usepackage{cancel}
\usepackage{mathrsfs}
\usepackage{amssymb}
\usepackage{amsfonts}
\usepackage{amsmath}
\usepackage{slashed}
\usepackage[font=small,labelfont=bf]{caption}

\newcommand \tr {\mbox{{\bf Tr}}}

\usepackage{graphicx}
\usepackage{bm}

\newcommand{\be}{\begin{equation}}
\newcommand{\ee}{\end{equation}}
\newcommand{\bes}{\begin{equation*}}
\newcommand{\ees}{\end{equation*}}
\newcommand{\bea}{\begin{eqnarray}}
\newcommand{\eea}{\end{eqnarray}}
\newcommand{\beas}{\begin{eqnarray*}}
\newcommand{\eeas}{\end{eqnarray*}}

\newcommand{\bmat}{\begin{bmatrix}}
\newcommand{\emat}{\end{bmatrix}}

\def\tr{{\rm tr}}
\def\Tr{{\rm Tr}}

\begin{document}

\numberwithin{equation}{section}
{
\begin{titlepage}
\begin{center}

\hfill \\
\hfill \\
\vskip 0.75in

{\Large \bf Exploring the Tensor Networks/AdS Correspondence}\\

\vskip 0.2in

{\large Arpan Bhattacharyya${}^{a,d}$, Zhe-Shen Gao${}^{a}$, Ling-Yan Hung${}^{a,b,c}$ and  Si-Nong Liu${}^{a}$  }

\vskip 0.3in

{\it ${}^{a}$ Department of Physics and Center for Field Theory and Particle Physics, Fudan University, \\
220 Handan Road, 200433 Shanghai, China} \vskip .5mm
{\it ${}^{b}$State Key Laboratory of Surface Physics and Department of Physics, Fudan University,
220 Handan Road, 200433 Shanghai, China}

{\it ${}^{c}$ Collaborative Innovation Center of Advanced  Microstructures,
Nanjing University,\\ Nanjing, 210093, China.}\vskip.5mm
{\it ${}^{d}$ Centre For High Energy Physics, Indian Institute of Science, 560012 Bangalore, India}

\end{center}


\begin{center} {\bf ABSTRACT } \end{center}
In this paper we study the recently proposed tensor networks/AdS correspondence. We found that the Coxeter group is a useful tool to describe tensor networks in a negatively curved space. Studying generic tensor network populated by perfect tensors, we find that the physical wave function generically do not admit any connected correlation functions of local operators. To remedy the problem, we assume that wavefunctions admitting such semi-classical gravitational interpretation are composed of tensors close to, but not exactly perfect tensors. Computing corrections to the connected two point correlation functions, we find that the leading contribution is given by structures related to geodesics connecting the operators inserted at the boundary physical dofs. Such considerations admit generalizations at least to three point functions. This is highly suggestive of the emergence of the analogues of Witten diagrams in the tensor network. The perturbations alone however do not give the right entanglement spectrum. Using the Coxeter construction, we also constructed the tensor network counterpart of the BTZ black hole, by orbifolding the discrete lattice on which the network resides. We found that the construction naturally reproduces some of the salient features of the BTZ black hole, such as the appearance of RT surfaces that could wrap the horizon, depending on the size of the entanglement region $A$.

\vfill

\noindent \today

\end{titlepage}
}

\newpage

\tableofcontents

\section{Introduction}

Recently there is a wave of new understanding of the AdS/CFT correspondence driven by developments in the understanding of quantum entanglement and its manifestation in a gravity theory via  the AdS/CFT correspondence. Most notably, the entanglement entropy of some region in configuration space in some states in a CFT corresponds to the area of a minimal surface in the bulk gravity theory \cite{Ryu,Ryu1}. It is by now clear that this minimal surface, often called the Ryu-Takayanagi (RT) surface, is profoundly connected to the Bekenstein-Hawking entropy formula. Till now, however, in the absence of a quantum gravitational theory, it is not clear how such a surface could emerge to encode the entanglement between microscopic degrees of freedom in the dual CFT. It was observed in \cite{Swingle} that such a connection between ``minimal surfaces" and entanglement entropy of a state does appear elsewhere, in a different context-- namely, in tensor network constructions of wavefunctions, a numerical technique that has been widely used by the condensed matter community. The network of tensors contracted with each other in specially chosen manner to suit the properties of the problem at hand -- whether it is close to the critical point of a phase for example-- appear to match some expected behaviour of a bulk gravitational theory. This inspired many subsequent works that attempt to make further concrete connection between the AdS/CFT and the tensor network. It is for example observed that the MERA tensor network, that originally inspired Swingle's observation \cite{Swingle}, is connected to the space of geodesics in AdS space, termed the kinematic space \cite{Czech,Czech1,Czech2,Czech3,Czech4}.\footnote{Interested readers are referred to  \cite{Orus} for a comprehensive review on tensor networks and its diverse applications in various  branches of theoretical physics. Also we list out some references \cite{references, references1, references2, references3} where the connection between reconstruction of  bulk spacetime and  entanglement in the context of holography have been explored. This list is by no means complete and the readers are encouraged to check the citations of these papers and the references within.}  More recently, \cite{HAPPY} proposes a concrete construction of a tensor network such that the entanglement entropy of the physical degrees of freedom is given precisely by the size of a minimal surface that cuts through the bulk of the tensor network. The important ingredient that goes into the proposal is perfect tensors, which have the magic property of being a unitary map between any choice of a set of half of the indices to the other half. This proposal is further developed in \cite{Qi1,Qi} so that there is a map between every state in the physical theory and a bulk tensor network, relaxing the restriction to the ``code-subspace" in \cite{HAPPY}.   Perfect tensors were shown \cite{Qi} to emerge rather typically for tensors with large bond dimension. Given this success however, there are obvious flaws to the proposal. To start with, it is known that the entanglement spectrum is always flat, which follows unfortunately also from the perfect-ness of the tensors. We would like to systematically study these tensor networks. A natural framework to discuss these tensor networks is the hyperbolic Coxeter group, which describes all tessellations in a  negatively curved space.  In section \ref{sec:introcoxeter}, we will begin with a brief review of the Coxeter group, and how it is related to a tensor network. In section \ref{sec:perfect}, we study properties of the perfect tensor network, and find that any tensor networks built from a collection of perfect tensors in a negatively curved lattice has no connected correlation functions between sufficiently local operators.
To remedy the problem, we explore the possibility that the wave-functions are built from tensors that depart from exact perfect tensors, albeit only by a ``small amount'', parameterized by a small number $\epsilon$. As we will describe in section \ref{sec:pert}, the leading order correction to the correlation function comes from a set of perturbations at nodes forming a connected geodesic connecting the operators acting at the boundary. This situation generalizes to three point functions.  It is tempting to interpret these as the analogy of Witten diagrams in the large mass limit, which emerges naturally as we perturb away from the perfect tensor background.  Then in section \ref{sec:btz}, we discuss how the BTZ black hole for example admits a tensor network construction by orbifolding the lattices, exactly as it is constructed by orbifolding in the continuous AdS space.  We will end with some concluding remarks in section \ref{sec:conclude}.  Various subtleties connected to a flat geometry and a discussion of a cure to the flat entanglement spectrum problem using ``weights'' are discussed in the appendix.

\section{Coxeter Group and Tensor Network}\label{sec:introcoxeter}

To understand tensor networks beyond the specific lattice structures suggested in \cite{HAPPY} and \cite{Qi}, we would like to make better contact between the isometries of the actual AdS space and the symmetries of the tensor network, and study systematically tensor networks that can fit into a negatively curved space. To do so, we will need to use the Coxeter group. 

\subsection{An overview of Coxeter group}
\begin{figure}
\begin{center}
\includegraphics[width = 0.46\textwidth]{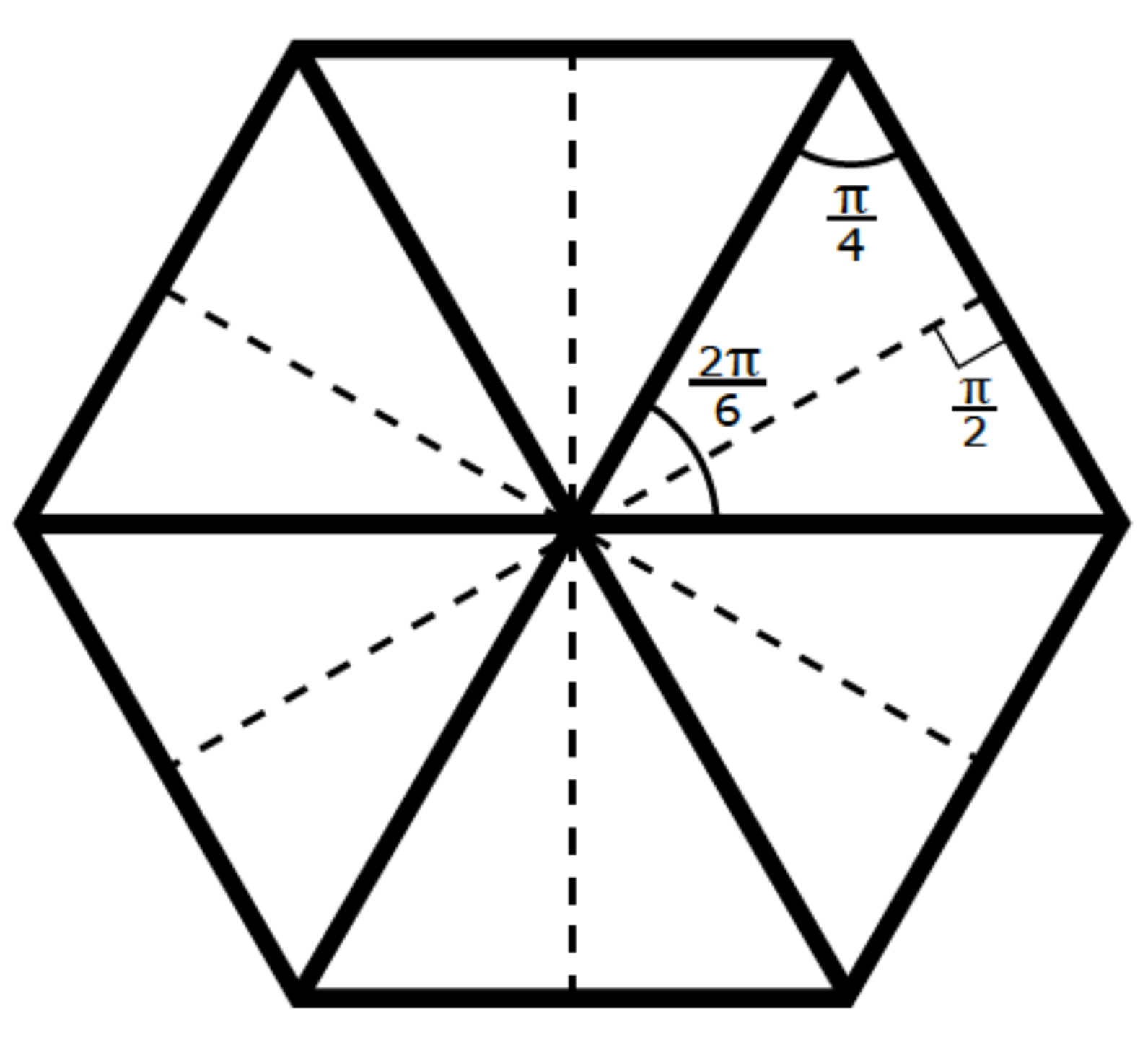}
\end{center}
\caption{This is an illustration of a tessellation specified by the triple $ [p=6, q=4]$.}
\label{fig:coxeter}
\end{figure}

The main idea we would like to explore is how one could systematically ``tessellate'' spacetime. That is achieved by exploiting reflections across various planes which in turn is connected to various properties of Coxeter group. First one can consider any Riemannian space  $M^{n}$ with constant curvature, i.e it can either be flat, positively curved or negatively curved.  Then consider polyhedra made up of locally finite number of intersections of half-spaces. We call them ``chambers'' or Coxeter polyhedra. We consider reflection across each of these faces of the polyhedra which are $(n-1)$ dimensions. Now consider another such polyhedra and reflections generated by their edges. Combining all these reflections we can ``tile'' the whole space. To be explicit, let us consider generating a picture like figure \ref{fig:hexacode} in a 2d space. One first determines the angles of a single triangle as will be described in the next section and figure \ref{fig:coxeter}. Then all the other triangles are related to the seed triangle by subsequent reflections across each of the sides of the seed triangle, and the process continues indefinitely. Reflection across each of the edge of the collection of all the triangles form the generators of the Coxeter group corresponding to this specific triangulation. i.e. The isometries generated by all these reflections form one particular ``Coxeter Group''.  We note that all isometries of the AdS space can be understood as combinations of reflections across codimension 1 planes. Therefore a specific triangulation based on the Coxeter group basically preserves a discrete group of the full isometries of the AdS space. 
This will be very crucial to our construction, as we will demonstrate that combination of these reflections would generate rotation (boost) and we will exploit that in the next sections.  The algebraic representation of these reflections across an edge on a Poincare disk will be described in detail in section \ref{sec:btzviacoxeter}. The group multiplication of these generators however can be readily recovered by simply looking at a picture of the triangulation, and combine series of reflections across edges of triangles. 
Suppose $F_{1},\cdots\,, F_{n}$ are the faces of the Coxeter polyhedra  characterized by the reflections  $s_{1}, \cdots\,, s_{n}$  satisfying $$s_{i}^2=1.$$ We also define the dihedral angle between the two face as $\frac{\pi}{m_{ij}}$  for all $m_{ij} \geq 2$ so they are always acute. Then  $$ (s_{i} \,s_{j})^{m_{ij}}=1.$$ Now these $m_{ij}$ are the elements of the {\it{Coxeter matrix}}.  We note that all the isometries of the hyperbolic space can be achieved by reflections in the flat embedding space. For detailed review of Coxeter group interested readers are referred to \cite{Coxeter,Coxeter1, Coxeter2,Wiki}.

\subsection{Towards constructing Tensor Network  }

For our case we will consider $H^2$ which will be the constant time slices of $AdS_{3}$. Now we want to tessellate this space time using the Coxeter reflections.  To tile a space by polytopes one important condition is that theses polytopes satisfy the Gauss-Bonnet theorem. In general because of this we can actually list all such polytopes when one tries to tessellate two sphere or flat (Euclidean) space. Now to tessellate $H^2$ we can start  with regular $p$-gons in hyperbolic 2-space with angles $\frac{2\pi}{q}$ . Any such polygon will tessellate $H^2$ with $q$ copies of these p-gons meeting at each vertex. Then we first divide these p-gons into p number of  triangles meeting at the center with the three angles $\frac{\pi}{q}\,,\frac{\pi}{q}\,,\frac{2\pi}{p}$.  Each of the angle of the polyhedra will be bisected by the sides of these triangles.  Now we divide further these triangles through the angle $\frac{2\pi}{p}$ into two right angled triangle there by bisecting one edge of the polyhedra also. Now these triangles will have angles $\frac{\pi}{q}\,,\frac{\pi}{2}\,,\frac{\pi}{p}.$ All these polyhedra used to tile the $H^2$ must satisfy the Gauss-Bonnet theorem implies that $$\frac{1}{p}+\frac{1}{q}+\frac{1}{2} < 1.$$  Figure \ref{fig:coxeter} is an  illustration of the case $[p=6,q=4]$.
We will use these triangles as the fundamental building blocks for constructing the lattice.  In the following we will describe a simple map between these tessellations and architecture of tensor networks.

\subsection{Some Recursion relations}

\begin{figure}
\begin{center}
\includegraphics[width = 0.46\textwidth]{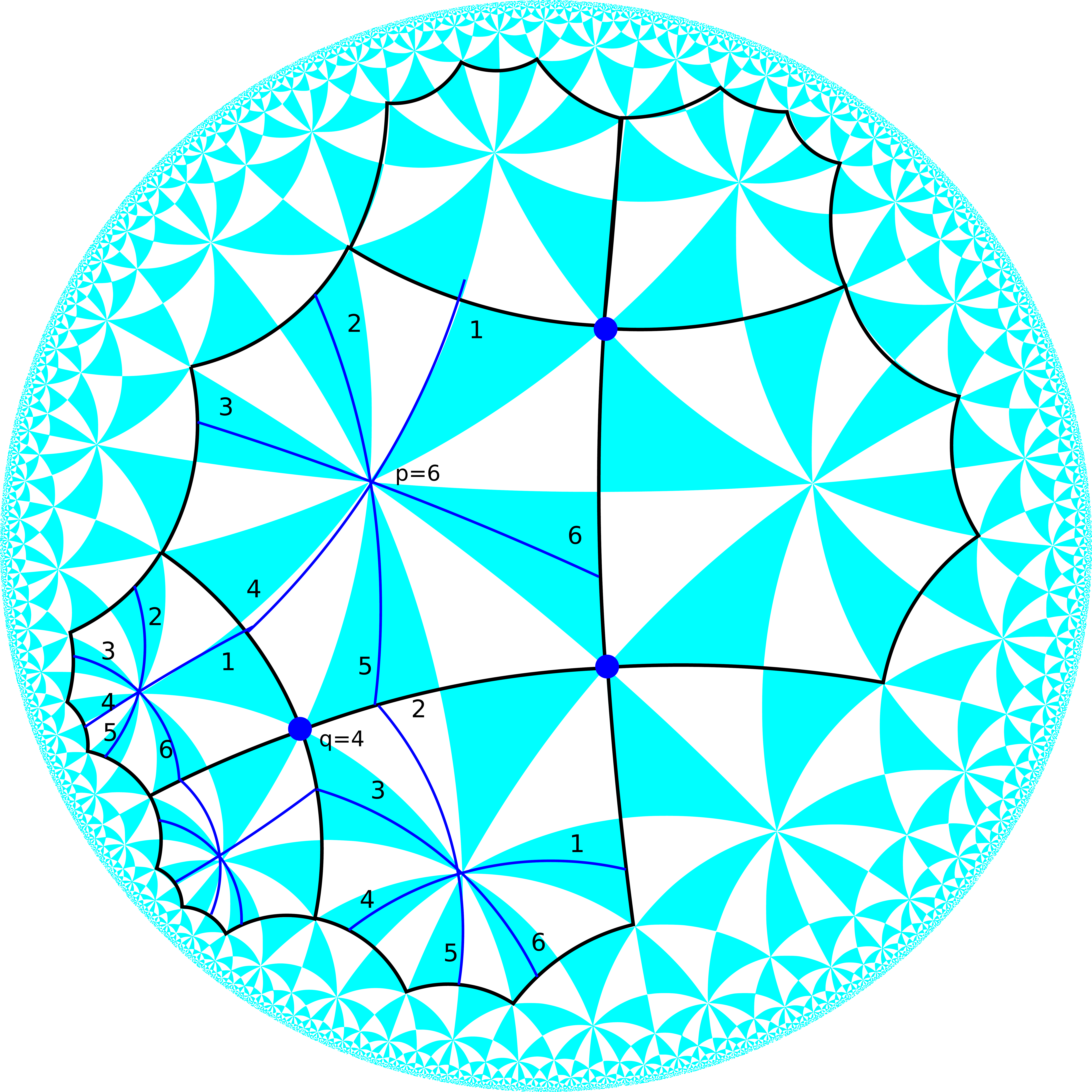}
\end{center}
\caption{This is an illustration of a tessellation specified by  $ [p=6, q=4]$. The numbers are labels of the 6 legs of the tensor. These tensors are generally not symmetric under permutation of the legs.}
\label{fig:hexacode}
\end{figure}

As demonstrated in \cite{HAPPY}, there is a recursion relation that can be derived, relating the number of tensors at each layer to the next. 
To define what we mean by a layer in the Coxeter tessellation of the hyperbolic space specified for example by the angles $[p,q,2]$  (if any of the entries takes value 2 it is usually omitted in standard notation which we will follow in all the subsequent sections.), where $p$ would be specifying the $p$-gon that holds a perfect tensor in its center, formed from $2p$ triangles, as already described. Each edge is shared by two $p$-gons , which should be interpreted as a tensor contraction between two tensors. One could see that the hexagon code in \cite{HAPPY} would correspond to a $[6,4]$ tessellation in this language. This is illustrated in figure \ref{fig:hexacode}.\footnote{Many of the Coxeter tessellation diagrams in this paper are based on the  diagrams obtained from Wikipedia \cite{Wiki}.} We note that conversely, providing only the data about the number of legs $p$ in each tensor and the number $q$ of nodes being adjacent to each other does not uniquely specify the tessellation. For example, the code in \cite{Qi} could have been equally well described by [5,5], although the map between vertices, edges and tensors and their contraction would have to follow some different rules. This is connected to the subtlety of what isometries are in fact preserved in the specific tessellation. We will in this paper base our discussion on the prescription as described above. 

To  assign a layer number to the nodes sitting at the center of a $p$-gon, we first pick a reference node in the bulk, and assign it a layer number $1$. Then the layer number of each neighbouring nodes sitting at the center of a $p$-gon sharing an edge with the reference $p$-gon increases by 1. Then the next layer is obtained by collecting all the $p$-gons sharing one edge with the layer -2 $p$-gons apart from the layer $1$ reference $p$-gon. This process can go on, giving a consistent assignment of layer number that increases as we move towards the boundary, as long as $q$ is even. (When $q$ is odd, there is a conflict in the assignment using the above simple rules. )

Using such an assignment, we will again come to the conclusion that no tensor is connected to other tensors in the same layer. Also, there would again be two types of tensors as in \cite{HAPPY}, one type where it has two legs connected to the previous layers, and the other type where there is only 1 leg connected to the previous layer.

\begin{figure}
\begin{center}
\includegraphics[width = 0.49\textwidth]{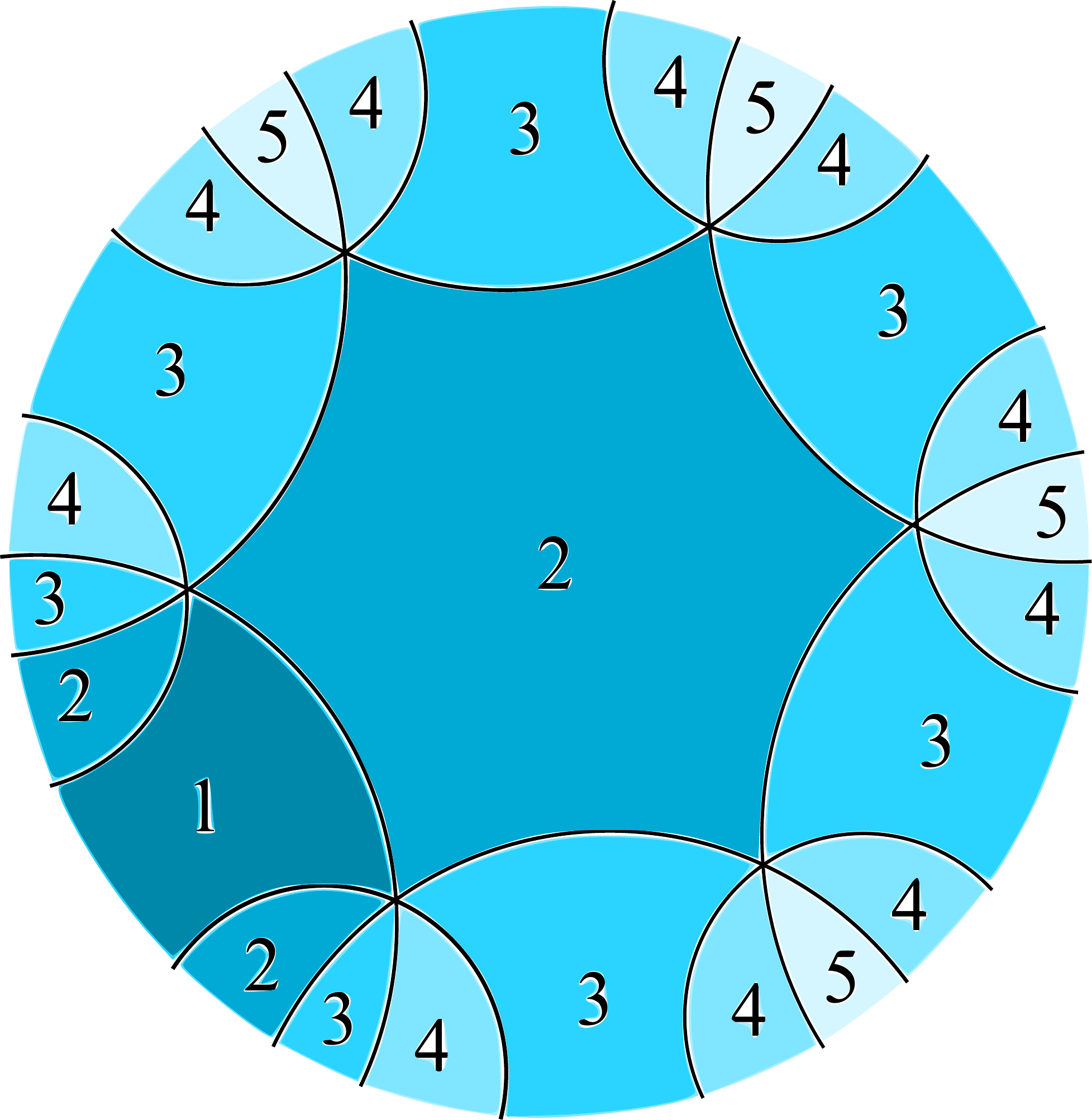}
\end{center}
\caption{In this picture, the numbers label the layer to which the hexagon belong to. The hexagons apart from 2 are truncated and we are only illustrating here the different tensor types we discussed in the text. It can be seen that at layer 2 there are 3 tensors of type $\tau_{2}(2)$ and at layer 3 there are 3 tensors of type $\tau_2(3)$ and 4 tensors of type $\tau_1(3)$. Finally there are two type $g(4)$ tensors, and 8 type $\tau_1(4)$ tensors.  }
\label{fig:recursion}
\end{figure}

At layer $n$, we denote type I tensors with two legs connected to the previous layer by $g(n)$.  Then, we can subdivide type II tensors in each layer with only 1 leg connecting to the previous layer into subsets $i$,  denoted $\tau_i(n)$, where $i$ denotes the shortest separation in layers they are from the next type I tensor. Clearly $1\leq i \leq q/2$, since $q$ $p$-gons meet at a vertex. 
The case in which $[p=6, q=4]$ is illustrated in figure \ref{fig:recursion}. 

One can see that the following recursion relation is satisfied:

\paragraph{$k=1$}\begin{enumerate}[(i)]
  \item$ g(n) \rightarrow (p-4)\tau(n+1)+g(n+1)$
  \item $\tau(n)\rightarrow g(n+1)+(p-3)\tau(n+1)$
\end{enumerate}

\paragraph{$k=2$}\begin{enumerate}[(i)]
  \item $g(n) \rightarrow (p-4)\tau_2(n+1)+2\tau_{1}(n+1)$
  \item $\tau_2(n)\rightarrow 2\tau_{1}(n+1)+(p-3)\tau_2(n+1)$
  \item $\tau_1(n)\rightarrow \frac{1}{2}g(n+1)+\tau_{1}(n+1)+(p-3)\tau_2(n+1)$
\end{enumerate}

\paragraph{$k\ge3$}\begin{enumerate}[(i)]
  \item $g(n) \rightarrow (p-4)\tau_k(n+1)+2\tau_{k-1}(n+1)$
  \item $\tau_k(n)\rightarrow 2\tau_{k-1}(n+1)+(p-3)\tau_k(n+1)$
  \item $\tau_i(n)\rightarrow \tau_{i-1}(n+1)+\tau_{k-1}(n+1)+(p-3)\tau_k(n+1)\qquad i=k-1,k-2\dots 2$
  \item $\tau_1(n)\rightarrow \frac{1}{2}g(n+1)+\tau_{k-1}(n+1)+(p-3)\tau_k(n+1)$
\end{enumerate}

One can check that this recovers the result in the appendix of \cite{HAPPY} where $p=5, q=4$  (the pentagon code).

The initial values are $N(\tau_k(1))=p$ and 0 for other kinds of tensors. And if we stop at some $n$ then we have \[N_{bounday}=N(g(n))+\sum_{i}N(\tau_i(n))\]and\[N_{bulk}=1+\sum_{n,i}(N(g(n))+N(\tau_i(n))),\]
where $N(X)$ denotes the number of tensors of type $X$. 
Some numerical results:
\begin{enumerate}
\item $\frac{N_{bulk}}{N_{boundary}}([6,8])\rightarrow 0.2508$
\item $\frac{N_{bulk}}{N_{boundary}}([6,10])\rightarrow 0.2504$
\item $\frac{N_{bulk}}{N_{boundary}}([6,16])\rightarrow 0.25$
\end{enumerate}

\section{Perfect Tensors} \label{sec:perfect}

In \cite{HAPPY}, it is demonstrated that a tensor network description of a wave-function such that it is defined on a negatively curved space and that each tensor is chosen to be a perfect tensor leads to a natural emergence of the RT surface when one computes the entanglement entropy of some connected group of chosen physical spins. Perfect tensors $T_{a_1\cdots a_{2n}}$ are tensors   with an even number of indices $2n$, such that an arbitrary separation of the indices into two groups $T_{I \alpha}$, $I = \{b_1,\cdots b_{n}\}$ and $\alpha = \{b_{n+1}\cdots b_{2n}\}$, $T_{I\alpha}$ is a unitary map satisfying
\be
T_{I \alpha} T^\dag_{\alpha I'} = \delta_{II'} = \prod_{i}^n\delta_{b_i b_i'}.
\ee

When the number of indices featuring in $I$ is less than that in the set $\alpha$, $T$ becomes a norm-preserving projector from $\alpha$ to $I$, thus satisfying
\be
T_{I \alpha} T^{\dag}_{\alpha I'} = \delta_{II'} D^s, 
\ee
where $s$ is the difference in the number of extra indices in the collection $\alpha$, and $D$ is the bond dimension of each index. 
The appearance of these delta functions play a significant role in the following discussions. It is behind the emergence of a bulk that appears local in many respect.

\subsection{Reduced density matrix and Modular Hamiltonian}
To begin with, we revisit the entanglement between physical degrees of freedom in some region $A$ described by this family of tensor network wavefunctions. As demonstrated in \cite{HAPPY}, a RT surface emerges, which specifies how a Schmidt-decomposition between region $A$ and its complement. 
Under the decomposition of the bulk wavefunction along the RT surface, the reduced density matrix is basically given by the isometry $V$ and it's complex conjugate,
\be
\rho_A = V_{x_1,x_2,\cdots x_{|RT|}}^{a_1,\cdots a_{|A|}} |a_1,\cdots a_{|A|}\rangle \langle a'_1,\cdots a'_{|A|}| V^{\dag\,\, a'_1,\cdots a'_{|A|}}_{x_1,x_2,\cdots x_{|RT|}}.
\ee

The modular Hamiltonian is thus a projector with eigenvalue $\log D$, where $D$ is the bond dimension, for the linear combo of states that mapped to the effective spins $x_i$, and infinity for any other linear combo.
The spectrum is thus completely flat. 


\subsubsection{Renyi entropy}

The fact that the entanglement Hamiltonian has a complete flat spectra also suggests that all the Renyi entropies for any Renyi index $n$ has exactly the same value. This is also observed in \cite{Qi}. We will explore the corrections to this issue.

\subsection{Stabilizers and symmetries}
Stabilizers correspond to symmetries. Any operator that do not commute with the stabilizers correspond to an operator charged under the symmetry.

Examples of perfect tensors were constructed from stabilizer code.
However, purely from the fact that these perfect tensors satisfy
\be
T. O = O' T
\ee
for arbitrary $O'$,  it means that
\be\label{stabilizer}
O^{'\,-1}.  T. O = T,
\ee
ie one can construct a set of operators that keep T invariant. In the case of spin 1/2 systems, where each index takes only 2 values,
it is clear that for $T$ having $L$- indices, we can construct a set of $2^L$ independent and commuting stabilizers, where $O$ can be chosen to be say $\sigma_z$  acting on each site and $O'$ obtained by conjugating O by $T$. This generates the complete set of stabilizers that specify the $L$-index tensor uniquely.
Therefore, every perfect tensor has a corresponding complete set of stabilizers. 

This construction can be generalized to any perfect tensor whose index takes $D$ values. We simply replace Pauli matrices by the generalized Pauli matrices. 
These are constructed by first defining
\be
X = \sum^D_k |k><k+1 (\textrm{mod} D)|, \qquad Z = \sum_k^D  \omega^{k-1 }|k><k|,
\ee
where $\omega = \exp(2\pi i/D)$. The rest of the $D^2$ basis, can be generated by products of $X$ and $Z$.

Given these symmetries at each node, which can be treated as global symmetries of the boundary theory,
one can expect that any correlation function is non-zero \emph{only if the operators have trivial charge}.
This is highly non-trivial given the amount of symmetry.

\subsection{Correlation functions}
We then turn our attention to correlation functions, which is a crucial aspect of the AdS/CFT correspondence \cite{Maldacena, Witten,Gubser1}. As we will see, a tensor network built from perfect tensors generically does not lead to any connected correlation function between local operators. 
In a discretized system, a local operator is one that acts on some number $P$ of spins, such that $P$ does not scale with the system size. 
The number of spins in the physical system grows exponentially, and the tensor network is a self-repeating structure. Therefore, we could presumably take  our boundary links as an effectively coarse grained set of degrees of freedom, and consider operators that act strictly locally on individual spins in the boundary layer.  This consideration can be relaxed as we generalize. 

\subsubsection{Operator pushing and stabilizers}
Action of an operator at the boundary can be readily pushed to action on the interior, i.e one needs to work out $O'$ using $O$. The perfect tensor at each node is constructed as a code stabilized by a set of stabilizers. Taking for example the hexagonal code in \cite{HAPPY} it is clear from  (\ref{stabilizer}) that $O^{'\, -1}$ and $O$ together form another stabilizer.
Therefore we can work backwards using the list of stabilizers and obtain $O'$ from any given $O$.

\subsubsection{Some examples- the Hexagonal code}

In this section we will demonstrate how correlation functions can be computed explicitly in the hexagon code. We will mainly consider the hexagonal holographic code consisting of six legs perfect tensors ($T_{\alpha\beta\gamma\delta\sigma\zeta}$)  constructed from 5 qubit stabilizer code discussed in \cite{HAPPY}. They are essentially characterized by the [6,4] triangulation which is already shown in figure \ref{fig:hexacode}. 
\par
\begin{figure}
\begin{center}
\includegraphics[width = 0.46\textwidth]{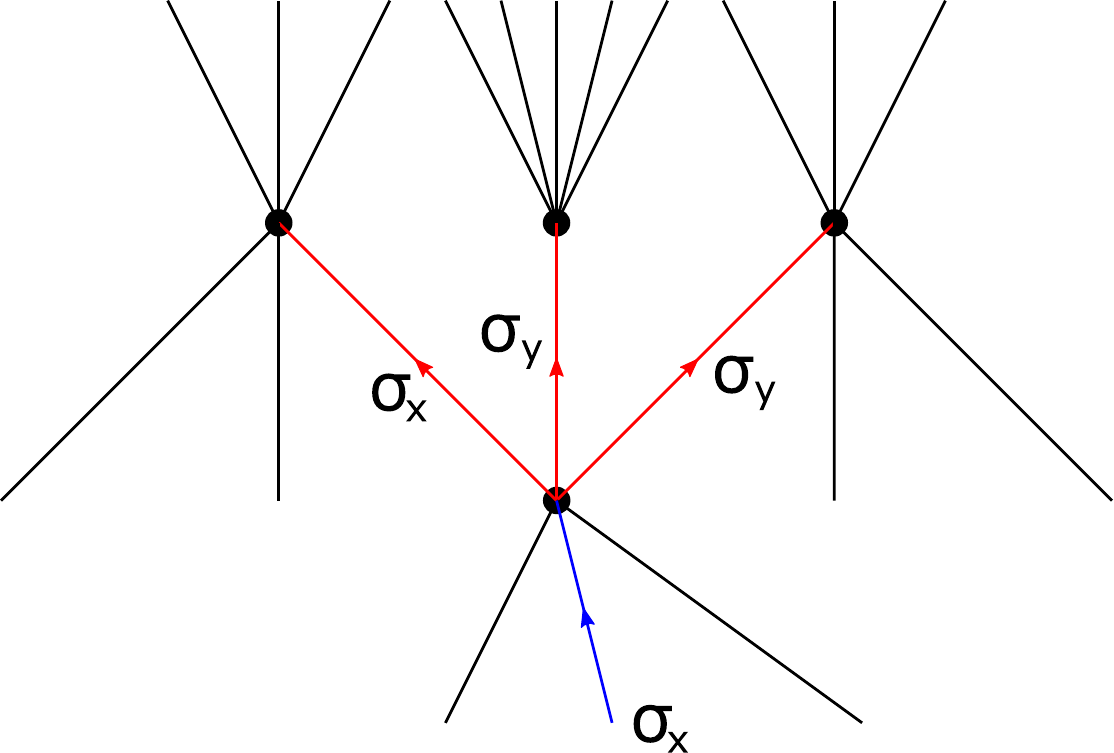}
\end{center}
\caption{This is an illustration of how the action of $\sigma_x$ on an external leg is replaced by products of other Pauli matrices in the interior leg in the hexagon code.}
\label{fig:push}
\end{figure}

 As we want to compute correlation functions of some operators $O_{i}$  we exploit the operator pushing property of the prefect  tensor as mentioned in (\ref{stabilizer}). The operators $O_{i}$ acts on the free legs of perfect tensors at the outer layer of the network. $O_{i}$'s are typically made up from Pauli operators so we need to know only the effect of pushing Pauli operators through three legs of the perfect tensors.  For example consider the six index perfect states $|\Psi>=T_{\alpha\beta\gamma\delta\sigma\zeta}|\alpha,\beta,\gamma,\delta,\sigma,\zeta>$.  $|\alpha,\beta,\gamma,\delta,\sigma,\zeta>$ is the orthonormal basis. The stabilizers for the tensor is given by \cite{HAPPY, Gottesman}
\bea
&& S_1 = XZZXII, \qquad S_2= IXZZXI, \qquad S_3 = IIXZZX,\qquad S_4 = XIIXZZ, \nonumber \\
&& S_5 = XXXXXX, \qquad S_6= ZZZZZZ,
\eea
where $X,Z$ are Pauli matrices $\sigma_{x,z}$ respectively.  Interested readers are referred to \cite{Gottesman} for more details about the stabilizer codes. 
One can show that by taking products of these stabilizers, it is possible to generate a new stabilizer such that one can make $X$ acts on any of the one spin, and that there are two $I$'s acting on another 2  arbitrary sites. For example, there is a stabilizer
\be
\tilde S |\Psi>=|\Psi>.
\ee
where,
\be\tilde S= \sigma_{x}\times I\times I \times \sigma_{x}\times \sigma_{y}\times \sigma_{y}.  \label{stabx}\ee
So from that it is evident that if we act on the  free legs of the tensor by $\sigma_{x}\times I\times I $ the effect is to act on the inner three legs by the operator $\sigma_{x}\times \sigma_{y}\times \sigma_{y}.$ Similarly for $\sigma_{y}\times I \times I$ one has to act with $\sigma_{y}\times \sigma_{z}\times \sigma_{z}$ on the inner three legs. This is illustrated in figure \ref{fig:push}.

   Now we will exploit this property to calculate the correlation functions. When the operator acts on the free legs after pushing them through a node it will act on the inner legs and hence will act on the bulk wavefunction formed by contractions of those inner legs. First we write down that bulk  wavefunction for this tensor network.  For now we focus on a single tensor.
\be
|\Psi>=T_{b_1 b_2 b_3 c_4 c_5 c_6}  U^{c_4 c_5 c_6}_{\cdots} |b_1,b_2,b_3 \cdots>  ,
\ee
where $U$ denotes all the other tensors $T$ and the boundary physical degrees of freedom are contracted to. 
$T_{b_1 b_2 b_3 b_4 b_5,b_6} $ is the perfect tensor  at the central node. Suppose we would like to compute
$|\tilde\Psi>= \sigma^{b_1}_x |\Psi>$. Using (\ref{stabx}) we immediately obtain
\be
 | \tilde\Psi>  = T_{b_1 b_2 b_3 c_4 c_5 c_6} \sigma^{c_4 c_4'}_{x}  \sigma^{c_5 c_5'}_{y} \sigma^{c_6 c_6'}_{y} U^{c_4' c_5' c_6'}_{\cdots} |b_1,b_2,b_3 \cdots> 
\ee

Note that $T$ can be viewed as a unitary transformation from $b_1,b_2,b_3$ to $c_4,c_5,c_6$. Therefore, what we have done is effectively obtaining a state
\be
 | \tilde\Psi> = \sigma^{c_4 c_4'}_x  \sigma^{c_5 c_5' }_y \sigma^{c_6 c_6'}_y  |c_4, c_5,c_6>    \otimes| c_4' c_5' c_6'  \cdots >,
\ee
where $U$ is also an isometry that defines a mapping from the boundary links to the links which the $\sigma_{x,y}$ are contracted with. 

 Now when we compute
\be
<\sigma^{x}>= <\Psi \tilde \Psi>.
\ee
This will give products of $\Tr\sigma^{x}$ and $\Tr\sigma^{y}$ which are zero. \\ 
Using this trick we can also check the  factorization of the correlator as described in the following section very easily. 
Most of the cases we will get zero connected correlations unless  we consider  correlations of highly non local operators -- namely those operators that act on close to more than half of the total number of boundary sites. 

\subsubsection{Proof of factorization of correlation functions}\label{proof}
When exactly do we get more non-trivial correlation functions?

One can see that the tricks of operator pushing would start to fail when we have too many operators acting on the boundary. The question is how many is deemed ``many"?

We can do the computation in a slightly different way. Decompose the bulk wavefunction into a product of two big tensors $U$ and $V$, each forming an isometry from some region $A$ (or its complement $A^c$ respectively) to a RT cut. Suppose the choice of region is one that contains the action of all the boundary operators.

Then the correlation function is
\be
\langle O \rangle= U^{a_1,\cdots a_{|A|}}_{x_1,\cdots x_{|RT|}} O_{a_1, a_1'} U^{{\dag}\,\, a_1'  \cdots a_{|A|}}_{y_1,\cdots y_{|RT|}} V^{b_a,\cdots c_{|A^c|}}_{x_1,\cdots x_{|RT|}} V^{\dag\,\,b_a,\cdots c_{|A^c|}}_{y_1,\cdots y_{|RT|}} 
\ee 
where $O$ denotes a group of operators acting on sites in region $A$.
The product of $V$'s and it's Hermitian conjugate generate a set of delta's on  $x$ and $y$. This is guaranteed when we place the tensors in a negatively curved lattice, in which the number of ingoing legs from the boundary to the interior layers keep decreasing exponentially, and that the tensors being perfect, are isometries from a layer to the next. 
We are thus left with the product of $U$'s with all the indices contracted with each other except for a few contracted to $O$, assuming that they are sets of local operators. 

Therefore, we are immediately left with
\be
\langle O \rangle = \tr O.
\ee
Moreover, since $O$ can be a set of local operators acting on multiple separated sites, we have
\be
\langle \prod_i O_i  \rangle = \prod_i \tr O_i.
\ee
We have a factorization of the operators.

\subsubsection{A remark on correlation functions and multipartite entanglement}
\begin{figure}
\begin{center}
\includegraphics[width = 0.5\textwidth]{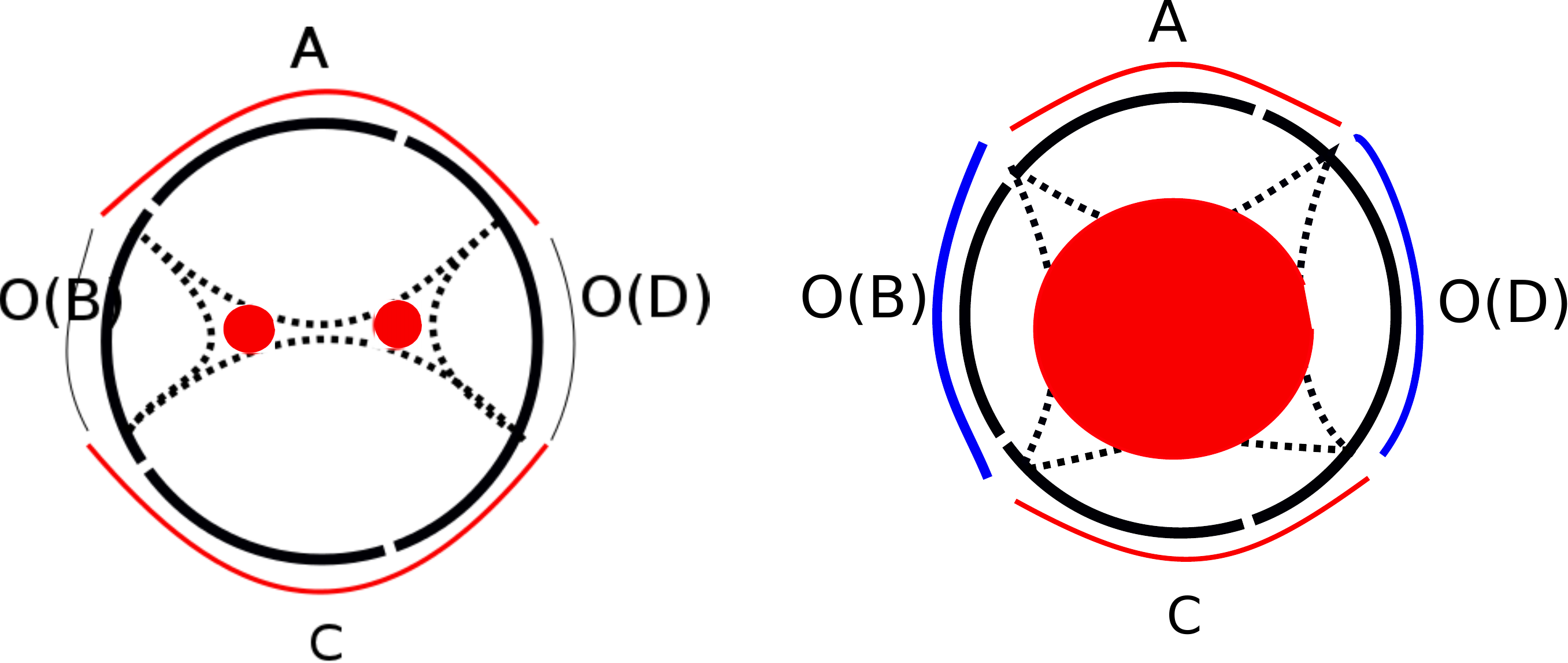}
\end{center}
\caption{We reproduce here the figure in \cite{HAPPY}. Operators with support in region $B$ and $D$ are inserted. The dotted lines correspond to geodesics forming the boundaries of the causal wedges of each boundary region. The regions colored red denote residual regions outside of the union of the four causal wedges. Connected correlation functions could feature when the residual region itself is connected.  }
\label{fig:GB}
\end{figure}

In \cite{HAPPY} there is a discussion on multi-partite entanglement. For a division of the boundary into 4 regions $A,B,C,D$, each of them is surrounded by an RT surface and the volume enclosed between the RT surface and the boundary region $A$ for example is denoted by the causal region of $A$ = $\mathcal{C}(A)$. It is noted that the union of $\mathcal{C}(A),\mathcal{C}(B),\mathcal{C}(C),\mathcal{C}(D)$ does not cover the entirety of the hyperbolic space and there are some residual regions. The size of the residual regions has a size that can be estimated from the Gauss-Bonnet theorem.  The figure describing these in \cite{HAPPY} is reproduced here in figure \ref{fig:GB} to illustrate its implications when we inspect correlation functions.

Suppose we consider correlation of operators with support in region $B$ and $D$.   The evaluation of the correlation functions $\langle O(B) O(D)\rangle$ involves tracing out, $A$ and $C$, which would lead to a set of delta functions pushed all the way to the geodesic which bounds $\mathcal{C}(A)$ and $\mathcal{C}(C)$. When $|A| |C| \gg |B| |D|$, it is clear that the residual region, denoted by the red dots in the figure, are disconnected. As a result, the correlation functions are actually disconnected since the delta functions would pinch off the throat. Only until $|A||D| \sim |B||D|,$  there is a connected region of residual regions, in which the correlation functions could be connected. But this immediately suggests that the support of the operator has to scale as system size before one obtains non-trivial connected correlation functions.

\section{Perturbations away from perfection}\label{sec:pert}

Since perfect tensors admit a lot of symmetries, as we demonstrated above, most of the correlation functions of local operators vanish. To obtain non-trivial correlation functions one has to move beyond perfect tensors. However, we would like to stay close to perfect tensors since it poses properties that can be identified with classical gravity, such as saturating the RT formula when computing entanglement entropy. We will therefore consider departure away from the perfect tensors as a small perturbation, and inspect their contributions to entanglement entropy and correlation functions.

The tensors at each node is thus given by
\be \label{perturb}
X = T + \epsilon\, t,
\ee
where $T$ is a perfect tensor, and $\epsilon$ is a small number. For example, in \cite{Qi} it is demonstrated that random tensors weighted by the Haar measure approach a perfect tensor in the large tensor dimension ($D$) limit. Therefore $\epsilon \sim 1/D$. 

The wavefunction is then given by
\be
|\psi\rangle = (\prod_{v} X_v)_{\alpha_1\cdots \alpha_L} |\alpha_1,\cdots \alpha_L\rangle
\ee
where $\prod_v X_v$  corresponds to the appropriate tensor contraction specified in a tensor network diagram, $X_v$ denote tensors at each node and $\alpha_i$ are boundary degrees of freedom. 

\subsection{Corrections to entanglement entropy}

One could compute the change to the entanglement entropy in the presence of perturbations $t$.
To leading order in $t$, the correction to the reduced density matrix involves only one $t$ located at some vertex in the tensor network.

\begin{figure}
\begin{center}
\includegraphics[width = 0.46\textwidth]{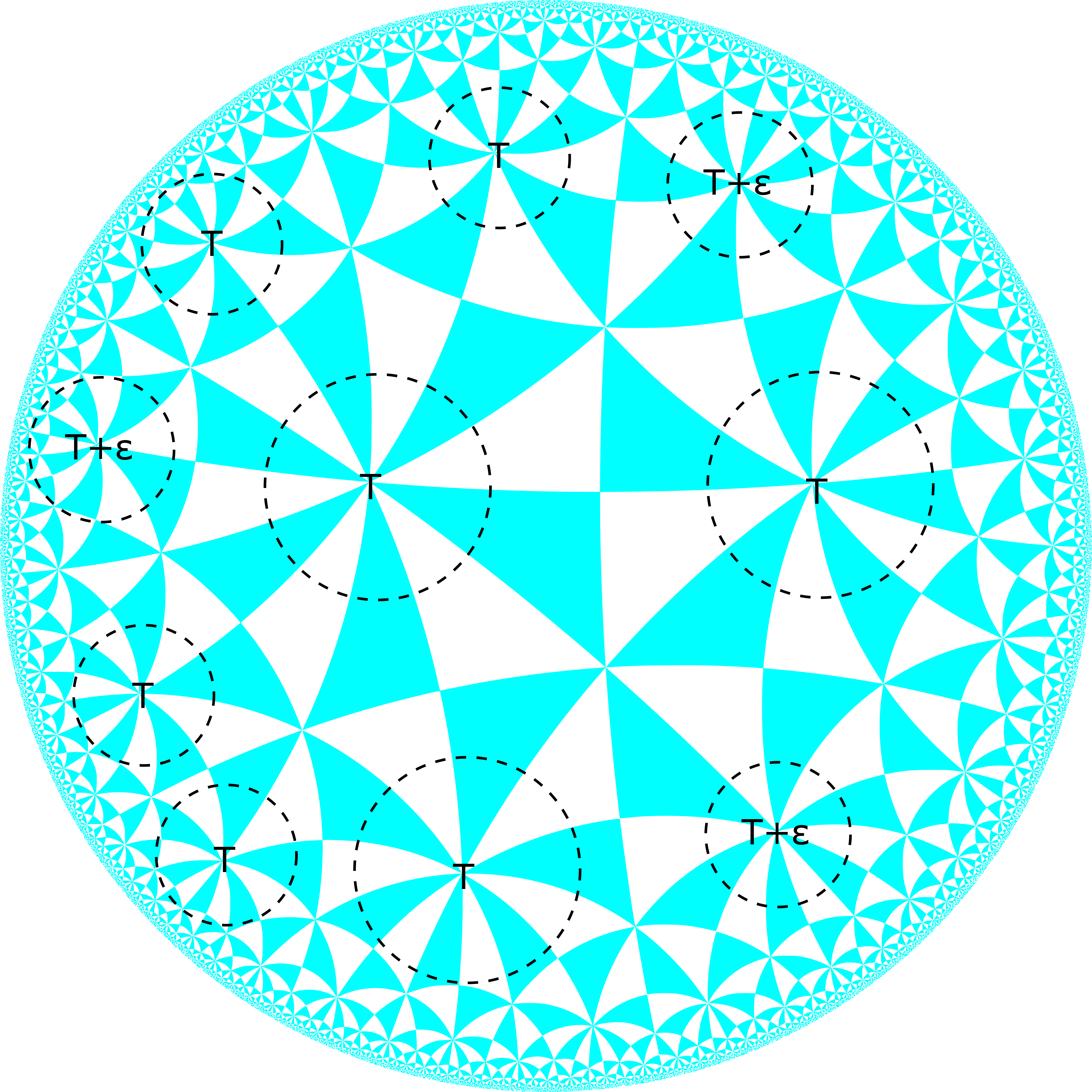}
\end{center}
\caption{In this figure, we would like to demonstrate that for sparsely  populated holes, one can show that they only change the normalization of the reduced density matrix, but do not contribute to changes in the entanglement spectrum.}
\label{fig:holes}
\end{figure}

Consider those cases in which this $t$ is located away from the RT surface.  In that case, in the computation of the reduced density matrix, one can consider splitting the calculation into three bits. Consider for simplicity the case of a single connected region $A$. Suppose $t$ is located ``outside'' of the RT surface. 
When computing the reduced density matrix, one first construct an isometry from the complement of $A$, denoted by $A^c$, to the boundary  plus a hole surrounding the node at which $t$ is located. For a negatively curved graph, this generically means we can draw a flow diagram from incoming boundary links in $A^c$ that flow out to the hole and the RT surface, defining an isometry from the boundary to the hole and the RT surface. 
Then when the links in $A^c$ are contracted, one obtains the identity operator for the out-going legs, which are legs at the hole and the RT surface. The identity operator is of course \emph{factorizable} and acts independently on the hole and the RT surface!. As a result, the spectrum of the reduced density matrix, which depends on the schimdt decomposition between $A^c$ and $A$ at the RT surface still has a flat spectrum as before dictated by the identity matrix there, except for an overall change of normalization $(1+ \epsilon\, \tr(T.t^\dag + t . T^\dag))$ that came from the hole.
 As a result, they do not contribute to any change in the entanglement entropy. This is illustrated in figure \ref{fig:holes}.
 We note here, however that negative curvature again plays a crucial role. 
Consider for example the four leg tensor considered in \cite{Qi}.  The tensor structure there is describable by the Coxeter group [4,5], if we inspect the arrangement of the self-repeating units. However, within each unit, the tensors are arranged in a way that is essentially flat. In that case, when a hole is dug within that unit, it is not necessarily possible to construct a flow diagram without internal loops such that one defines an isometry from the incoming legs near the boundary to the legs connecting to the next layer and to the hole.  We will discuss this issue in more detail in the appendix.

A similar calculation can be done when $t$ is located ``within'' the RT surface. In that case, the trace over region $A^c$ leads again to a simple delta function on the links of the RT surface, and that the reduced density matrix is basically 
\bea
\delta \rho_A &=& \epsilon\, T_{x_1,x_2,x_3,..., x_{L}} U^{\beta_1\,\cdots \beta_N\, x_1,x_2,x_3,... x_{L-1}}_{\alpha_1\cdots \alpha_A} |\alpha_1 \cdots \alpha_A\rangle \langle \alpha_1'\cdots \alpha_A'| U^{\dag \,\, \beta_1\,\cdots \beta_N\, x_1',x_2',x_3',..., x_{L-1}'} t^\dag_{x_1' x_2' x_3' ... x_{L-1}'}  \nonumber \\
&&+ (T \leftrightarrow t))
+ \textrm{h.c.}, 
\eea
where now $U$ is an isometry constructed from the boundary to the hole and the RT surface, and we are working with an $L$-leg tensor $T$. For precisely the same reason as the previous scenario, the presence of $t$ in the interior do not lead to any change of the entanglement spectrum, and so the entanglement entropy is not altered to this order of the perturbation.

In fact, for that reason, any disconnected appearance of $t$, small number of them relative to bulk size, away from the RT surface, could only contribute to change in the overall normalization, and no change to the entanglement spectrum, and thus preserve the entanglement entropy.

The leading order change has to come from $t$ that falls \emph{exactly} on top of the RT surface, which then actually changes the spectra of Schmidt coefficients along the RT surface.
In fact the change to the reduced density matrix due to perturbation $t$ at site $i$ on the RT surface is given by
\bea
\delta \rho_A (i) &=& \sum_{i}^\gamma U_{\alpha_1\cdots \alpha_A}^{\beta_1 \beta_{i-1}\,\cdots x_1 x_2 x_{L-1}  \beta_{i+1}\cdots\beta_N} T_{x_1,x_2,x_3,\beta_i} 
U^{\dag\,\,{\beta_1 \beta_{i-1}\,\cdots x_1 x_2 ..x_{L-1}  \beta_{i+1}\cdots\beta_N}}_{\alpha_1' \cdots \alpha'_A} t^\dag_{x_1,x_2,x_{L-1},\beta_i}  \nonumber \\ &&|\alpha_1\cdots \alpha_A\rangle \langle \alpha'_1 \cdots \alpha'_A|  + (T\leftrightarrow t) + \textrm{h.c.}.
\eea

This is highly reminiscence of the bulk calculation of changes to the entanglement entropy, in which the entanglement entropy is only sensitive to the changes in the metric on top of the RT surface. 

To linear order, the correction to the entanglement entropy thus comes from
\be
S + \delta S = -\tr ((\rho^0_A+ \delta \rho_A) \log (\rho_A^0+ \delta \rho_A) ).
\ee

The form of $\rho_A$ is a projector, in which it has equal eigenvalues along some $|D|^\gamma$ dimensional sub-space $H_\gamma$, and vanishing eigenvalues in the orthogonal $D^{|A|}- D^\gamma$ dimensional  subspace $H_{\bar{\gamma}}$. On the other hand, $\delta \rho_A  $ generically has off-diagonal elements $\delta \rho_{A\,\,\gamma \bar{\gamma}}$ and also elements purely in the orthogonal subspace $\delta \rho_{A\,\, \bar{\gamma}\bar{\gamma}}$.

Diagonalizing $\rho_A^0 + \delta \rho_A$, to linear order in the small perturbation, it is forced to have eigenvalues of the form
\be
U(\rho^0_A + \delta \rho_A)U^\dagger= \frac{1}{D^\gamma}\textrm{diag} \{1 + \epsilon_1,\cdots, 1+ \epsilon_{D^\gamma}, \epsilon_{D^\gamma+1} ,\cdots \epsilon_{D^|A|} \},
\ee  
where
\be
\sum_i^{D^{|A|}} \epsilon_i =0.
\ee
The leading contribution to $\delta \rho_A$ contains only $t$ very close to the RT surface. In practice therefore the number of eigenvalues $\epsilon_i$ in the orthogonal subspace is actually less than $\gamma D^{L-1}$, where $L$ is the number of legs of the tensor.

Therefore, we have
\be
\delta S = \frac{1}{D^\gamma} \sum^{D^\gamma}_i \epsilon_i (\ln D^\gamma -1)  - \sum_i^{D^{|A|}-D^\gamma}  \frac{\epsilon_{D^\gamma + i}}{D^\gamma} \ln  (\frac{\epsilon_{D^\gamma + i}}{D^\gamma}).
\ee
There are two issues of note. First, there is necessarily an $\epsilon \ln \epsilon$ contribution, which is non-perturbative in the small parameter, whenever $\delta \rho_A$ leads to extra eigenvectors in the orthogonal subspace.  If the holographic code is generically an isometry from boundary degrees of freedom to bulk degrees of freedom, even with more interesting entanglement spectrum than the flat spectrum that follows from perfect tensors, one would expect that any perturbation that involves the orthogonal subspace could lead to such non-perturbative terms. It would be a very strong test of the proposal to understand if such a term could arise in the AdS/CFT correspondence. We note that for generic perturbations of couplings these would not arise. The only possibility is perhaps when a coupling perturbation is mixed with the change of cut-off scales.  

A second issue of note is that if $\epsilon_j =0$ for $ j> D^\gamma$, then $\delta S =0$ following from $\tr \delta \rho_A=0$. This is expected because within the subspace $V^\gamma$ entanglement is maximal. 
To obtain non-trivial perturbation to the entanglement entropy it is necessary to move on to a second order perturbation.

\subsubsection{Quadratic order correction}
To complete our discussion, it is instructive to inspect also second order contribution. 
To second order in $t$, the density matrix takes the following form
\be
\rho = \mathcal{N}^2(|\psi_0\rangle \langle \psi_0| + |\psi_0\rangle \langle \psi_1| + \textrm{h.c. } + |\psi_0\rangle \langle \psi_2| +  \textrm{h.c. } +|\psi_1\rangle \langle \psi_1 |\cdots)
\ee
where $|\psi_n\rangle$ denotes corrections to the state corresponding to the sum of all configurations in which $n$ of the nodes is replaced by $t$, and that $\mathcal{N}$ is the normalization, which itself admits an expansion in $t$. 
For later convenience, let us denote
\be
P_n = \tr_{\bar{A}}\sum_{i+j=n} |\psi_i\rangle \langle \psi_j |.
\ee
 Precisely for the same reason as already discussed, when one replaces two nodes by $t$ away from the RT surface, their contributions to the Renyi entropy are canceled by the normalization. Let us focus on the case in which $\delta \rho_A$ projects onto the same sub-space as $\rho_A$, so that  non-perturbative logarithmic terms do not appear.
In that case the leading quadratic contribution to the Renyi entropy comes from $|\psi_1\rangle \langle \psi_1 |$. In particular, since $t$ aligns in the same subspace that the perfect tensors $T$ projects into,  $ |\psi_0\rangle \langle \psi_2| +  \textrm{h.c. } $ do not contribute to the quadratic order in entanglement entropy.

The systematics of  $\mathcal{N}$ in this case is as follows:
\be
\mathcal{N}^2 = \frac{1}{D^\gamma(1+ \epsilon \alpha +\epsilon \beta/D)},
\ee
where 
\be
\alpha = \sum_{i\notin \gamma} \tr(T^\dag_i. t_i + t^\dag_i . T_i) , \qquad \beta = \sum_{i\in \gamma} \tr(T_i^\dag.t_i + t_i^\dag. T_i ).
\ee
where the $\tr$ means that all four indices between $T$ and $t$ are contracted with each other.
We then have
\be
\tr(\rho_A^n) = \frac{1}{D^{n \gamma}}(1- \alpha -\beta/D)^n \bigg(\tr \mathbb{I}^n + n \tr(\mathbb{I}^{n-1}( P_1+ P_2 + \cdots))  + \frac{n(n-1)}{2}  \tr\mathbb{I}^{n-2} P_1^2  + \cdots\bigg).
\ee
Focussing on the quadratic contribution, we have
\be
\tr P_1^2 = \sum_{i\in \gamma} \tr[(T_i^\dag .t_i + t^\dag_i . T_i)^2] + \sum_{i,j\in \gamma,\, i \neq j} \tr(T_i^\dag .t_i + t^\dag_i . T_i)\tr(T_j^\dag .t_j + t^\dag_j . T_j) + \sum_{i,j\notin \gamma} \tr(T_i^\dag .t_i + t^\dag_i . T_i)\tr(T_j^\dag .t_j + t^\dag_j . T_j).
\ee
The last two terms are canceled out by the normalization. We finally have
\be
S_n = \frac{1}{1-n} \ln \tr(\rho_A^n) = \gamma \ln D - \epsilon^2 \frac{n}{2D} \sum_{i\in\gamma} \tr[(T_i^\dag.t_i + t^\dag_i. T_i)^2].
\ee

In the following, we will discuss how $t$ enters into the computation of the correlation function. Looking ahead, comparing with (\ref{opdim1}, \ref{opdim2}) the above quadratic correction to the Renyi entropy can be related to the conformal dimension of operators as
\be
\epsilon^2 \sum_{i\in\gamma} \tr[(T_i^\dag.t_i + t^\dag_i. T_i)^2] = \gamma  \sum_{ab}\epsilon^2 \lambda_{ab}^2= \gamma \sum_{ab} \exp(-2\Delta_{ab}).
\ee

\subsection{Correction to correlation functions}

Given (\ref{perturb})
\be\label{pertOO}
\langle \psi | O_1(i) O_2(j)|\psi \rangle = (\prod_v X^\dag_v)_{\alpha_1 \cdots \alpha_i \cdots \alpha_j \cdots } 
O_1(i)_{\alpha_i \alpha_i'}  O_2(j)_{\alpha_j \alpha_j'} (\prod_p X_p)_{\alpha_1 \cdots \alpha_i' \cdots \alpha_j' \cdots } 
\ee
where $O_m(i)$ is some operator located at site $i$.

Now consider the correction to linear order in $\epsilon$. This means that in (\ref{pertOO}) exactly one node features some tensor $t_v$, while all other nodes are occupied by perfect tensor $T_v$. Given the property of the perfect tensors, the contraction of $\alpha_s$ where $s\neq i$ and $s\neq j$ means that most of these tensors $X_v, X_p$ contract to give delta functions. Exactly as in the previous section, the correlation function would factorize into
\be
\langle \psi | O_1(i) O_2(j)|\psi \rangle  = \tr O_1(i) \tr O_2(j) \tr (T. t)
\ee

\begin{figure}
\begin{center}
\includegraphics[width = 0.5\textwidth]{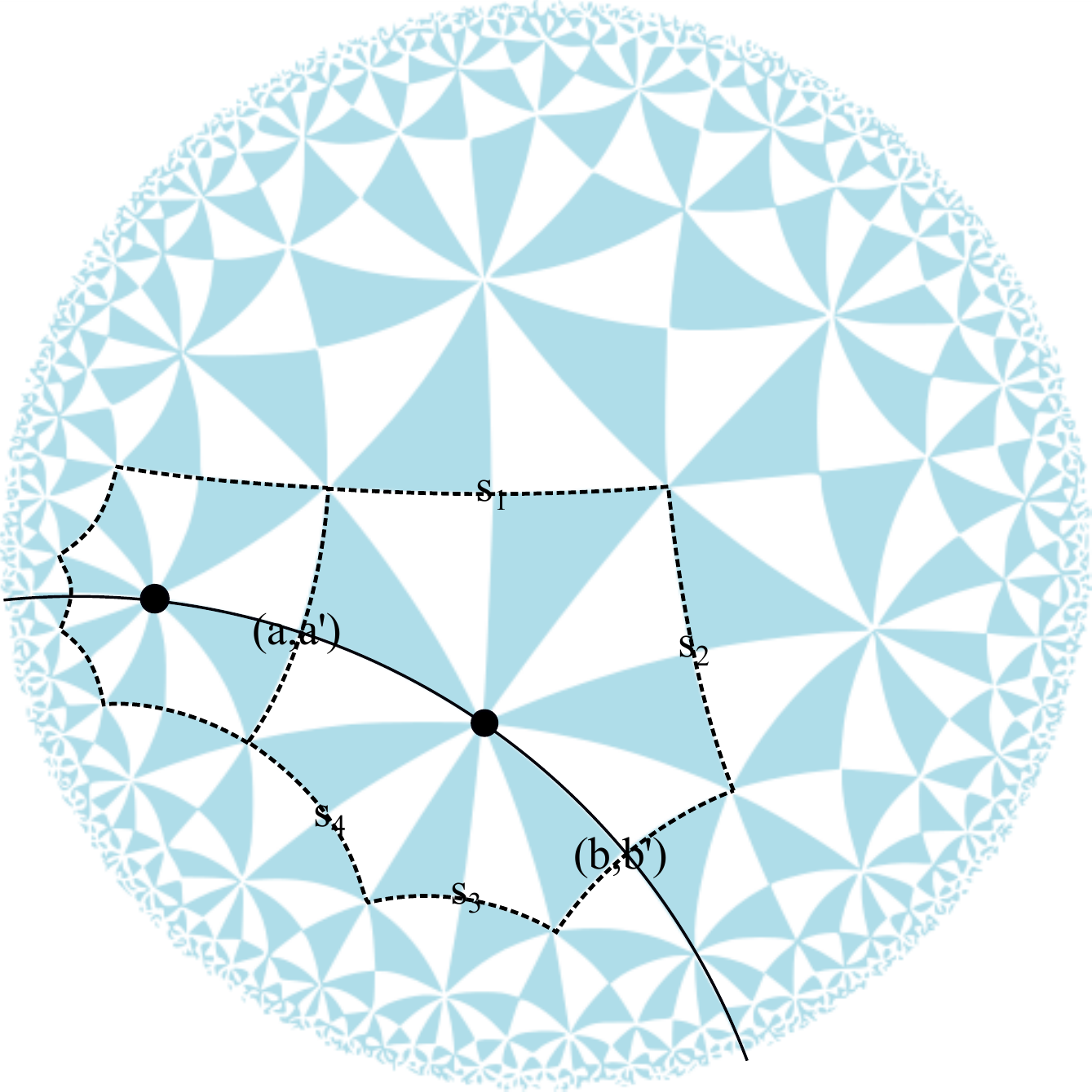}
\end{center}
\caption{The picture illustrates a two point function whose leading contribution follows from a set of perturbations $t$ along nodes along a geodesic connecting the operators inserted at the end points. The legs of a tensor is labeled according to the contraction defined in the transfer matrix (\ref{opdim1}). }
\label{fig:2pt}
\end{figure}

To obtain a connected correlation function, it is thus clear that we need a string of nodes replaced by a departure from perfect tensors. To obtain the leading term in the $\epsilon$ expansion, it is thus given by a string of nodes that connects $O_1(i)$ and $O_2(j)$ along a \emph{geodesic}, the shortest path through the tensor network that connects the two boundary sites $i$ and $j$ as illustrated in figure \ref{fig:2pt}.

In fact the leading contribution would take the form
\be 
\langle \psi | O_1(i) O_2(j)|\psi \rangle = \epsilon^{|\gamma|}  \tr(\prod_{v}^{|\gamma|} T_v  O_1(i) O_2(j) \prod_p^{|\gamma|}t^\dag_v + \cdots) = \epsilon^{|\gamma|}\tr( O_1(i) O_2(j)\prod_v  \mathcal{T}_v)
\ee
where $|\gamma|$ is the number of nodes along the \emph{geodesic}, and $\cdots$ denote the terms where the $T$'s and $t^\dag$ at arbitrary node $v$ along the geodesic are interchanged. This is in turn rewritten in terms of a transfer matrix $\mathcal{T}$. Let us illustrate this with the hexagon code which is characterized by the [6,4] lattice. This is illustrated in figure \ref{fig:2pt}.

For the orientation of the tensor as indicated in the diagram, 
\be \label{opdim1}
(\mathcal{T}_{o_1})^{ab}_{a'b'} = [(T. t^\dag + t.T^\dag)_v]^{ab}_{a'b'} = T_{a s_1 s_2 b s_3 s_4} t^\dag_{a' s_1  s_2 b' s_3 s_4} + \textrm{h.c.}.
\ee
Here, the $T.t^\dag$ denotes contraction of only 4 of the 6 indices of each of these tensors, and the sub-script $o_1$ denote the transfer matrix for a specific
orientation of the geodesic.  
Suppose $T$ and $t$ are  same at every node for a geometry  that is homogenous, then the correlation function would be controlled by the eigenvalues of the transfer matrix $\mathcal{T}$.

In fact, suppose in the 6-leg network constructed in \cite{HAPPY} the transfer matrix is diagonalized to
\be \label{opdim2}
(U^{-1}\mathcal{T}U)^{ab}_{a'b'} = \lambda_{ab}\delta_{a a'} \delta_{b b'},
\ee
for some appropriate unitary matrices $U$, then the 2 point correlation functions are given by
\be
\langle O_{ab}(i) O_{ab}(j)\rangle = \epsilon^{|\gamma|} \lambda_{ab}^{|\gamma|} \delta_{12},
\ee
where $O_{ab}$ is assumed to be operators in the appropriate basis that diagonalizes $\mathcal{T}$. These operators would correspond to conformal primaries, as pointed out in \cite{Qi}.
The conformal dimension are given by
\be
\Delta_{ab} = - \log( \epsilon\, \lambda_{ab} ),
\ee
for a holographic code such that $|\gamma| = \log \Delta x$, where $\Delta x$ is the distance between two sites $i,j$ at the boundary connected by the geodesic $\gamma$. That $\epsilon\, \lambda$ is small for the approximation to be valid means that we are naturally working with an operator with large conformal dimension. This coincides with the AdS intuition that the Green's function of a very massive particle is approximated by geodesics. 

We note that this approach also naturally recovers the counting in \cite{Qi} in which correlation functions are found to scale also with $1/D^{| \gamma |}$, since $\epsilon$ plays the role of $1/D$ when one averages over random matrices of large dimensions $D$. 

Before we move on, let us also comment on the calculation here and that of the entanglement entropy in the previous section. It is well known that in a 1+1 d CFT, the Renyi entropy can be computed by considering essentially correlation functions of local twist operators, whose dimension depends on the central charge and Renyi index $n$ by
\be
\Delta_n = \frac{c}{6}(n- \frac{1}{n}).
\ee
A computation of the correlation function would require that a non-trivial result appears at order $\epsilon^\gamma$, where $\gamma$ is the length of the geodesic, leading to an entanglement entropy $S_{EE}\sim \gamma \log \epsilon $.
The lion's share of the entanglement entropy however starts with order $\epsilon^0$. This entanglement follows from correlation between highly non-local operators. The computation of the reduced density matrix in the previous section recovers small corrections of this large amount of entanglement, starting with $\epsilon \log \epsilon$. This puts in tension between the relationship of the tensor network and the CFT. We note that indeed as observed in the previous section, such perturbations away from perfection falls short of recovering the $n$-dependence of the Renyi entropy of a CFT. It is expected  that other fixes, such as the introduction of weights, is necessary to correct this problem. It is also possible that there is a non-trivial interplay between computing the $n\to 1$ limit in which the conformal dimension necessarily becomes small, and the small $\epsilon$ expansion.

As a constraint of preserving isometry, one would expect that if the geodesic is oriented in a different direction, the spectra of operators remain the same.
If such is the case, we should require that another transfer matrix corresponding to a different orientation in the six leg tensor network
\be
(\mathcal{T}_{o_2} )^{ab}_{a'b'} = T_{s_1\, a \,  s_2 s_3\, b\, s_4 } t^\dag_{s_1 a'  s_2 s_3 b' s_4} + \textrm{h.c.}
\ee
to have the same spectra as $\mathcal{T}_{o_1}$. 

The symmetry of this tensor naturally allows for one more orientation of the geodesic
\be
(\mathcal{T}_{o_2} )^{ab}_{a'b'} = T_{s_1 s_2 a s_3  s_4 b} t^\dag_{s_1   s_2 a' s_3  s_4 b'} + \textrm{h.c.}
\ee

Now geodesics can be orientated in very different ways. For an [L,M] lattice, geodesics are most naturally running across the L-gon in $L/2$ different ways. (see for example figure \ref{fig:3pt}, in which the three geodesics meeting at a point happen to have three different orientations. ) It appears that for an $L$ leg tensor embedded in a [L,M] lattice it is natural to have $L/2$ different orientations of the geodesic that respect lattice symmetry, and that we should require that they preserve the same spectrum to allow for the same spectra of operators independently of the relative orientation of the inserted operators. 

\subsection{Three point correlation function and fusion coefficients}

We can continue our quest and move on to three point correlation functions. Following the same logic,
we can see that leading connected three point correlation functions are given by products of perturbations $t$ away from perfect tensors $T$ along three paths that connect the three points $x_i, i=\{1,2,3\}$ where operators are inserted at the boundary, and that these paths are joined at a point $z$. This configuration must be such that the number of $t$'s involved is minimal, and thus the paths connecting $z$ and $x_i$ should be a boundary to bulk geodesic. Moreover, the point $z$ should be chosen such that the sum of these three geodesics have a minimal length among all these geodesics.

In a discrete graph, it is not immediately obvious where the point $z$ is located. Let us resort to the  knowledge of continuous AdS space, which would be a reasonable approximation when the lattice representation of the AdS space is sufficiently fine-grained.

It is already observed in \cite{Janik} that the three point correlation functions of some massive spin states can be approximated in the bulk by three geodesics joining at a point. The correct central point that minimizes the length of this path was also found there, using AdS isometry.

Explicitly, it is found that the 3-point correlation function takes the following form
\be \label{AdS3pt}
\langle O_{\Delta_1}(x_1) O_{\Delta_2}(x_2) O_{\Delta_3}(x_3)\rangle = \exp(W[1,2,3]), \qquad W = \sum_i \Delta_i \ln \frac{\epsilon}{z^2 + (x-x_i)^2},
\ee
where $\epsilon$ is the radial $z$ cutoff surface, and $(x,z)$ is the position of the bulk point at which the three boundary-bulk geodesics join.

It is then found that a suitable choice of $x$ and $z$ gives the right lengths of these geodesics such that
\be
\langle O_{\Delta_1}(x_1) O_{\Delta_2}(x_2) O_{\Delta_3}(x_3)\rangle \sim \frac{1}{|x_{12}|^{\Delta_1+\Delta_2-\Delta_3}|x_{13}|^{\Delta_2+\Delta_3-\Delta_1}|x_{23}|^{\Delta_2+\Delta_3-\Delta_1}},
\ee 
precisely the correct universal  form of the three point correlation function of a CFT.

Putting this result in our context, we have obtained a set of ``primary operators'' in the previous subsection, by diagonalizing the \emph{transfer matrices}.
Therefore working in that basis, the three boundary to bulk geodesics would automatically generate the expression (\ref{AdS3pt}), from which the knowledge of the existence of an appropriate bulk junction leads to a correct form of the three point function.

This then gives us a handle of how the \emph{fusion coefficients} between operators are computed.

At precisely the junction, the transfer matrices along the three boundary-to-bulk geodesics would be contracted with a tensor at the junction. Taking again the 6-leg perfect tensor construction, the three point function takes the form
\bea \label{fusionF}
\langle O_1(1) O_2(2) O_3(3) \rangle &=& O_{\Delta_1\, a_1,b_1}(( \mathcal{T}_1)^\gamma_1)^{a_1 b_1}_{a'_1 b'_1} O_{\Delta_2\, a_2,b_2} ( (\mathcal{T}_2)^{\gamma_2})^{a_2 b_2}_{a'_2 b'_2}  O_{\Delta_3\, a_3,b_3}  ((\mathcal{T}_3)^{\gamma_3})^{a_3 b_3}_{a'_3 b'_3} \nonumber \\
&&   \times (T_{a'_1 s_1 a'_2 s_2 a'_3 s_3} t^\dag_{b'_1s_1 b'_2  s_2 b'_3 s_3 }+ t_{a'_1 s_1 a'_2 s_2 a'_3 s_3} T^\dag_{b'_1s_1 b'_2  s_2 b'_3 s_3 }).
\eea

This is illustrated in the following figure \ref{fig:3pt}.
\begin{figure}
\begin{center}
\includegraphics[width = 0.5\textwidth]{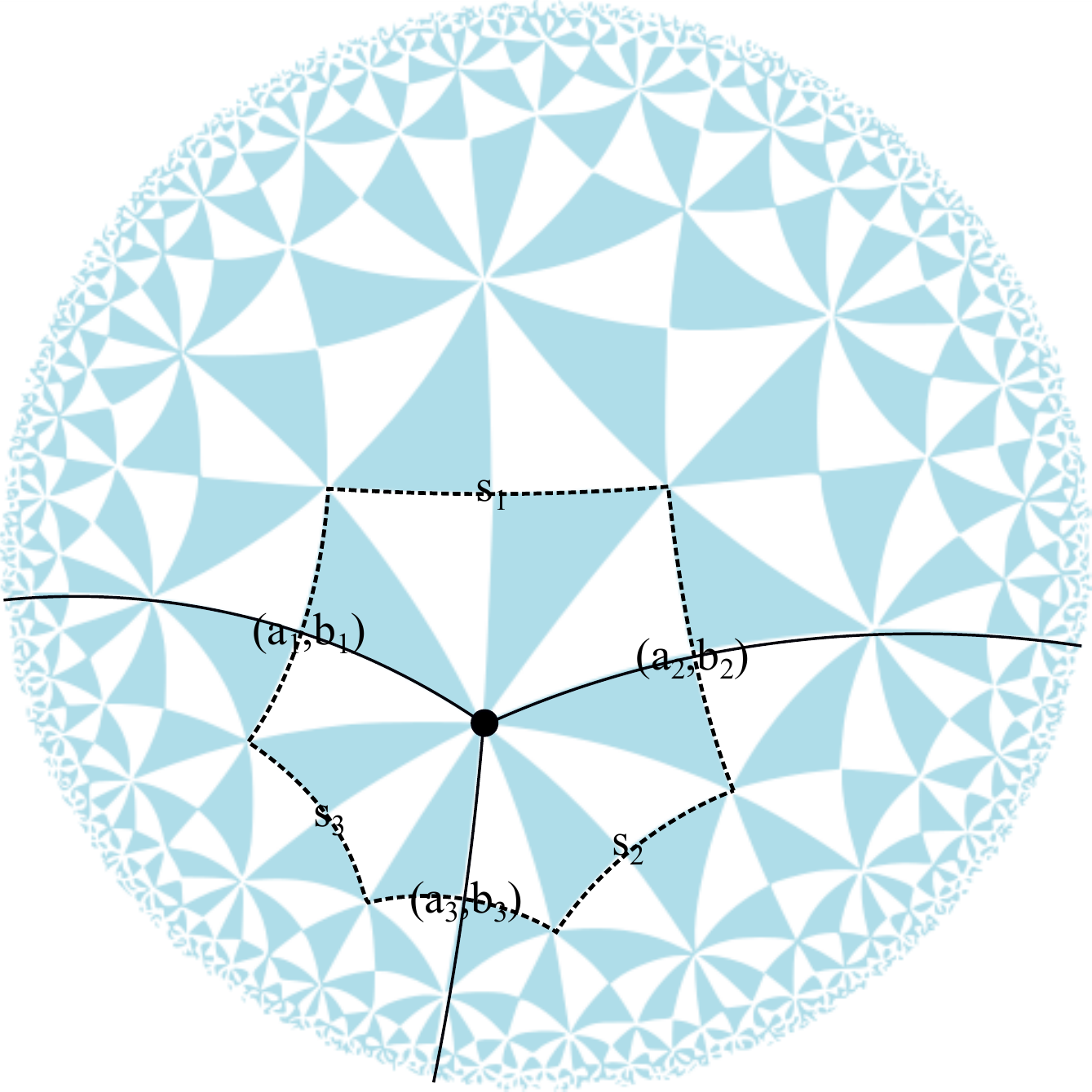}
\end{center}
\caption{The picture illustrates a 3 point function whose leading contribution involves three geodesics meeting at a point. The tensor at which they meet defines a fusion matrix, and the legs of the tensor are labeled according to (\ref{fusionF}). }
\label{fig:3pt}
\end{figure}

One can immediately see that the \emph{fusion matrix} is given by
\be
\mathcal{F}^{(a_1,b_1),(a_2,b_2),(a_3,b_3)} =   (T_{a_1 s_1 a_2 s_2 a_3 s_3} t^\dag_{b_1s_1 b_2  s_2 b_3 s_3 }+ t_{a_1 s_1 a_2 s_2 a_3 s_3} T^\dag_{b_1s_1 b_2  s_2 b_3 s_3 }).
\ee

One important requirement following from isometry, is thus that the fusion coefficients should have at least the same eigenvalues when any of the four indices of $T$ is contracted with the corresponding index in $t$, since depending on the specific diagram, the junction can have different orientation and it is expected that the fusion coefficient should stay ``unchanged'' (possibly up to a change of basis at each boundary site).

\subsection{Constraints }

From rotational invariance  of two point function which also follows from the homogeneity and isotropy of the tensor network, it is natural to require that
\be \label{const}
\mathcal{T}_{o_i}= \mathcal{T}_{o_j}.
\ee 
A generic complex perturbation $t$ have $D^L$ number of components for an $L$ leg tensor at each node. The condition (\ref{const}) gives $2 (L/2-1) \times D^{4}$ constraint equations. Now from the three point function analysis of the previous section, we come to know there will be $C^L_3$ different orientations of $\mathcal{F}_i.$ e.g. for the six leg tensor, $\mathcal{F}_i$ are constructed by picking any three of the indices out of the 6, and contract them between $T$ and $t^\dag$: 
\be
\mathcal{F}^{1 (a_1,b_1),(a_2,b_2),(a_3,b_3)}_{1} =   (T_{a_1 a_2 a_3 m_4 m_5 m_6} t^\dag_{b_1 b_2 b_3 m_4 m_5 m_6} + t_{a_1 a_2 a_3 m_4 m_5 m_6} T^\dag_{b_1 b_2 b_3 m_4 m_5 m_6}),
\ee
and so on.

If we demand rotational invariance coming from homogeneity and isotropy, most naively, we should set
\be
\mathcal{F}_i=\mathcal{F}_j.
\ee
The number of equations is equal to $2 (C^L_{3}-1)D^6$

It is evident that with $L<6$, it becomes difficult to satisfy all the constraints barring isolated solutions that may or may not appear.  We tested these conditions with the 4-leg tensor and the 6-leg tensor detailed in \cite{HAPPY}, and found that in the case of the 4-leg tensor, it is possible to solve the 2-pt function constraints together with 2 of the 3 constraints following from the 3-pt correlation. In the case of the hexagon code we could solve all the 2-pt constraints together with 9 of the 19 constraints in the 3 point function, using only real matrices $t$, which appears already to be doing far better than the naive counting would suggest. 

Another possibility is that the constraints that we are imposing is too stringent. After all, the basis operators acting on different sites need not coincide precisely, and that perhaps it is reasonable to require only that the eigenvalues of the transfer matrices match. 
\be
\mathcal{T}_{o_i} = U_{ij} \mathcal{T}_{o_j}V^{-1}_{ij},
\ee
where $U, V$ also need not be identical, since our transfer matrices are not Hermitian upon complex conjugating, which exchanges the pair $(a,b)$ and $(a',b')$. Rather, it is ``Hermitian" upon exchange of $(a,a') \leftrightarrow (b,b')$. Matching only eigenvalues (or strictly speaking singular values for asymmetric matrices) leads to $(L/2-1) D^2$ equations. Once these $U_{ij},V_{ij}$ are determined, they have to be recycled in the computation of the 3-point function. This reduction of constraints therefore is not significant enough to avoid anisotropy of 3-pt fusion. It is perhaps more natural that a generic tensor network with a proper gravitational interpretation should admit some fairly large number of legs at each node.

\section{Simulating BTZ geometry in the Tensor Network}\label{sec:btz}
Having a concrete understanding of the subset of isometries of the AdS space preserved in a given lattice using the Coxeter group, we are now in a position to make better use of these symmetries and study the BTZ black hole, which is obtained by orbifolding the AdS$_3$ space. 

\subsection{Orbifolding and BTZ}
In the previous sections we have demonstrated that using the tensor network we can compute 2 pt and 3pt correlation functions which give us the necessary information about conformal dimensions and  fusion matrices. It is also shown that they have the correct scaling properties as expected from CFT. So at least these tensor networks reproduce  some basic features of AdS/CFT although we are still far from  giving it a proper gravity description.  But nonetheless all these observation suggest that one can implement these tensor networks to understand dynamics of the gravity and possibly how Einstein equation emerges. In this section we will take another important step towards understanding these concept by constructing a BTZ black hole with this network purely using the symmetry considerations. We will see that our construction gives all the information which has been known from these tensor networks in the previous studies \cite{HAPPY} but our construction is more physical and it backs up our idea of constructing tensor network using the Coxeter group constructions. \\
We first review the construction of BTZ black hole by orbifolidng Poincare AdS following \cite{Carlip, MS}. Below we quote the Euclidean BTZ metric,
\be 
ds^2=\frac{(r^2-r_{+}^2)(r^2-r_{-}^2)}{r^2\,l^2}d\tau^2+\frac{r^2\,l^2}{(r^2-r_{+}^2)(r^2-r_{-}^2)}dr^2+(d\phi+\frac{r_{+}(i r_{-})}{r^2}d\tau).
\ee
$r_{+}$ and $r_{-}$ denote the location of the two horizons. $\tau$ is the Euclidean time.  We know that Riemann curvature tensor takes the same form  for the $AdS_{3}$ and BTZ black hole. We now define the following coordinate transformations,
\be \label{trans}
w=\Big(\frac{r^2-r_{+}^2}{r^2-r_{-}^2}\Big)^{1/2}\exp\big\{\frac{r_{+}+r_{-}}{l}(\phi+i\,\tau)\big\}\,, \quad z=\Big(\frac{r^2-r_{+}^2}{r^2-r_{-}^2}\Big)^{1/2}\exp\big\{\frac{r_{+}}{l}\phi+\frac{i\,r_{-}}{l}\tau\big\}.
\ee
Using this we get,
\be
ds^2=\frac{l^2}{z^2}(dw d\bar w+ dz^2).
\ee
and $z>0.$ So this restrict the space time in the upper half plane and can be thought of quotienting the three hyperbola. Now the periodicity condition on $\phi\sim \phi + 2\pi $ implies,
\be \label{pl4}
w\sim e^{\frac{4\pi^2}{\beta}}w, \quad z\sim e^{\frac{4\pi^2}{\beta}} z ,
\ee
where in the case at hand we have restricted our attention to non-rotating black holes, and thus real temperature $\beta$. The orbifolding transformation thus correspond to an overall re-scaling of $\omega$ and $z$. 
Now for our network construction we generally consider constant $\tau$ slice of $H_{3}$ and so we will have $w=\bar w$, and that there is no angular momentum, so that the temperature $\beta$ is real, and $r_- =0$.  We obtain an orbifold of the $AdS_{3}$ geometry by identifying points in Poincare AdS related by the symmetry transformation to obtain the BTZ geometry. This is the crucial ingredient in constructing the corresponding tensor network. 
\subsection{BTZ via Coxeter group} \label{sec:btzviacoxeter}
As demonstrated earlier we know that using the generators of Coxeter group which are basically some reflections about a chosen plane one can construct all the Lorentz generators.  So combination of a finite number of reflections can generate any Lorentz transformations. The mirrors appropriate for the specific reflection can be found by constructing planes in the flat embedding space and then obtain their  intersection with the hyperbolic space. Intersections of planes (that pass through the origin) with an $H_2$  centered also at the origin are geodesics. Boosts are therefore generated by reflections across geodesics in $H_2$. Orbifolding an $H_2$ tessellation respecting these reflection symmetries determine the BTZ geometry on a constant time slice.   To demonstrate this we write the equation of the $H_2$ space in embedding coordinate \cite{MAGOO},
\be \label{hyperbola}
X_{0}^2+X_{3}^2-X_{2}^2-X_{1}^2=R^2
\ee
$R$ is the radius of the hyperbola.  To go to the Poincare coordinate we will use,
\be
X_{0}=\frac{z}{2}\Big(1+\frac{R^2+w\bar w}{z^2}\Big)\,,X^1=\frac{R}{z}(w+\bar w)\,,X^{2}=\frac{z}{2}\Big(1-\frac{(R^2-w\bar w)}{z^2}\Big)\,, X_{3}=\frac{R}{2\,i\,z}(w-\bar w).
\ee

In the following, we will restrict ourselves to the $t=0$ slice, so that $X_3=0$, and that the slice, an AdS$_2$ space, is invariant under the orbifolding transformation (\ref{pl4}). 
Meanwhile, $X_{0}$ and $X_{2}$ transforms and the other two remains unchanged. From the perspectives of these flat space embedding coordinates, the orbifold transformation is in fact a Lorentz boost in the $X_0-X_2$ plane.

Writing,
\be \label{pl2}
\tilde X_{0}= X_{0} \cosh(\eta)+X_{2} \sinh(\eta)\,, \quad \tilde X_{2}= X_{2} \cosh(\eta)+ X_{0}\sinh(\eta).
\ee
one immediately see that
\be
\eta = \frac{4\pi^2}{\beta}.
\ee

As described already in the introduction of the Coxeter group, every isometry of the hyperbolic space can be recovered by a reflection across planes also embedded in flat space. Therefore, the orbifolding transformation of a given point in the hyperbolic space, which corresponds to a boost in embedding space, can be recovered by specifying a plane across which the point reflects.
Under reflection across a plane defined by $\vec{X}.\vec{n}=0$, with $\vec{n}$ specifying the normal vector of the plane, a point transforms as
\be \label{pl1}
 \vec{\tilde X}=\vec{X}-2\, \Big(\vec{X}.\vec{n}\Big) \vec{n}.
\ee

\begin{figure}
\centerline{
 \includegraphics[height=6.5cm]{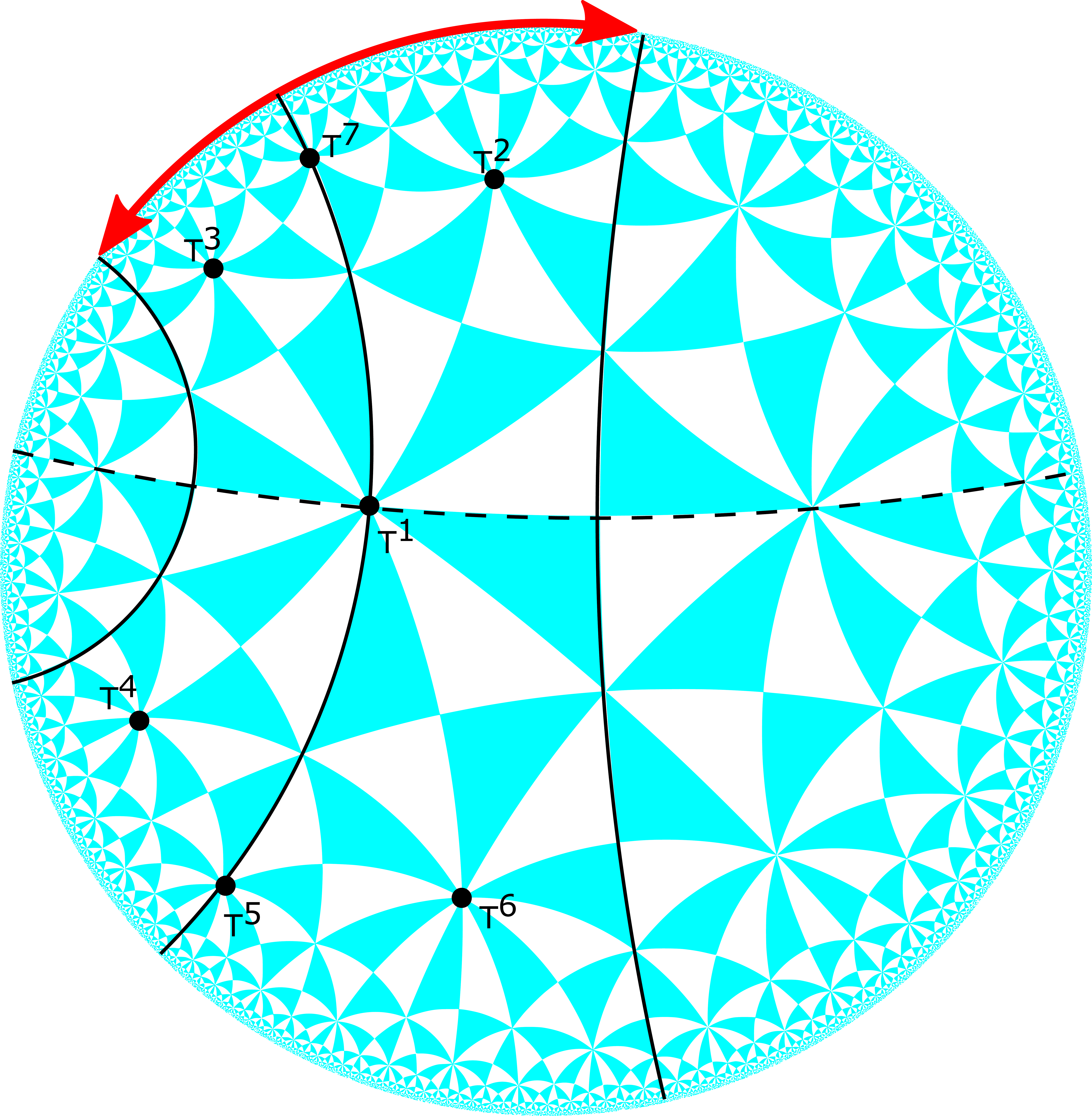}
\hskip 2.cm \includegraphics[height=6.5cm]{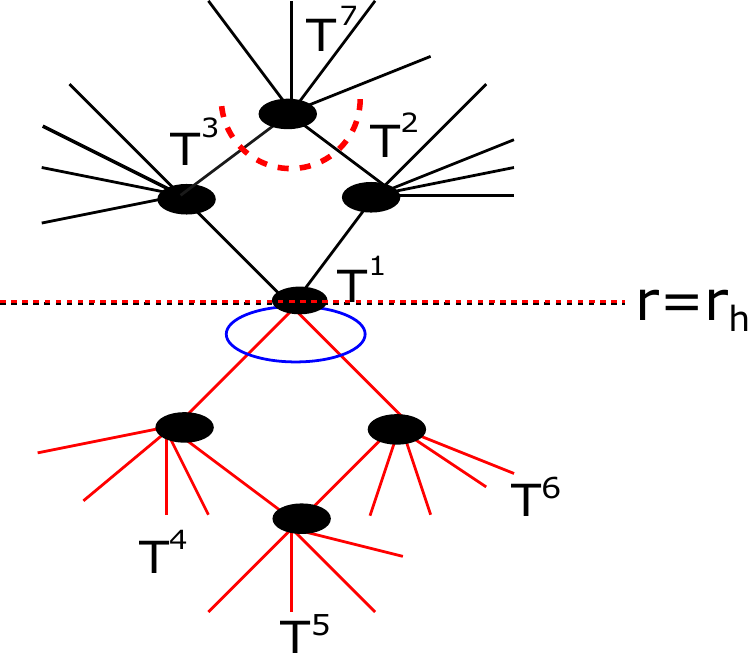}~~~~~~
}

\caption{On the left panel, we illustrate the orbifolding that is done on the [6,4] tessellation. The black curves are the mirrors generating the desired boosts. The fundamental region we have chosen is indicated by the red arrow.  The dashed line is the apparent horizon, across which degrees of freedom are entangled. The nodes corresponding to different tensors are marked. The tensor network that follows is illustrated on the right panel.  Also we have marked the two RT surfaces corresponding to tracing out node 7 and node 2 and 3 in red.
}
\label{fig:btz}
\end{figure}

Comparing (\ref{pl2}) and (\ref{pl1}), we get a consistency condition on the normal vector $\vec{n} =(n_0,n_1,n_2,n_3)$
\be
n_0^2- n_2^2 = -1
\ee
Now without loss of generality, we would first start with a reference mirror, specified by $n_0=0$ and thus $n_2=1$. Points acquiring a boost (\ref{pl1}) under reflection across this plane lie on a plane satisfying
\be
\frac{X_2}{X_0} = - \tanh \frac{\eta}{2}. 
\ee
Starting with this plane and reference mirror, we would like to obtain subsequent mirrors so that the images acquire further boosts again given by (\ref{pl2}) . It is not hard to see that the next mirror which would produce the correct boost on the image produced by the reference mirror is related to the reference mirror by precisely the same boost (\ref{pl2}). Since $n_0=0$ for the reference mirror, it is basically located at $X_2=0$. Under the boost (\ref{pl2}) it means that the next mirror satisfies
\be\label{mirror1}
\frac{X_2}{X_0} = \tanh \eta.
\ee
In fact all subsequent set of mirrors that generate boosts by reflection on the previous images are all generated by the reference mirror by sequentially  boosting it by (\ref{pl2}).  We can plot these planes on the Poincare disk. This is done by solving for the intersection of these planes with the hyperbolic space:
\be \label{pl3}
\vec{X}.\vec{n_{i}}=0\,, \quad X_{0}^2+X_{3}^2-X_{2}^2-X_{1}^2=R^2,
\ee
where $\vec{n}_i$ is the normal of one of these infinite sequence of planes. 
These intersections are geodesics on the hyperbolic space.  One can plot them on the Poincare unit disk $(Y_1,Y_2), \,\, Y_1^2 + Y_2^2 \leq 1$ via the map
\be
Y_1 = \frac{X_2}{1+ X_0},\qquad Y_2 = \frac{X_1}{1+ X_0} .
\ee The reference mirror we have chosen is thus a vertical line at $Y_1=0$ at the center of the Poincare disk. Subsequent curves can be obtained readily. For example, the second mirror on the hyperbolic plane corresponding a curve satisfying both (\ref{hyperbola}) and (\ref{mirror1}) is given by the parametric equation:
\be
Y_1 =  \frac{X_0\tanh\eta}{1+X_0}, \qquad Y_2 = \frac{\sqrt{X_0^2 - \cosh^2\eta}}{\cosh\eta (X_0+1)}. 
\ee
The collection of these mirrors in the Poincare disk looks like a set of ``parallel curves'' that do not meet.   To obtain an orbifolded tensor network, it now corresponds to first picking a tessellation that contains the set of mirrors generating the orbifolding boosts, and then identify all the nodes related to each other by reflection by any of these mirrors. This would lead to periodic contractions of legs. The tensor network would acquire the topology of a cylinder. We take the [6,4] lattice as an example and illustrate it in figure \ref{fig:btz}.  In the next section we will compute explicitly the entanglement entropy that follows from this tensor network.

\subsection{Entanglement Entropy and RT surface in the BTZ background}
In this section we will compute the entanglement entropy for the particular tensor network  and we will demonstrate that the result matches with BTZ result perfectly.  To do so, we first briefly  review the $AdS_{3}$ calculation of the entanglement entropy for BTZ using the standard RT formula which can be found in \cite{Ryu1} details. We will just point out some key points which we will use in this section. As for the BTZ as the starting point is not a pure state so entropy for the region $A$ and its complement $A^{c}$  are not equal.\\
The main conclusion is that we will have generally two different geodesics - 1.) One geodesic corresponding to $A$ do not cross the region and 2.) Other one  corresponding to $A^{c}$ which wraps the black hole one time.  Next we will determine these two geodesics from the tensor network. 
This is illustrated in the figure \ref{fig:btz}.
 Now we write down a specific wavefunction for this fundamental domain below as illustrated in figure \ref{fig:btz}.
\begin{align}
\begin{split} \label{wave1}
 |\Psi>&= \Big(T^{7}_{abc_1c_2c_3c_4}T^{2}_{e a c_5 c_6 c_7 c_8 } T^{3}_{e_1 b c_9  c_{10} c_{11} c_{12} } T^{1}_{e\, t_1 e_1 t_1 f_1 g_1} \\&T^{4}_{f_1 x  c_{13}c_{14}c_{15}c_{16} } T^{6}_{g_1  x_{1} c_{17}c_{18}c_{19}c_{20} }T^{5}_{x\,x_{1} c_{21}c_{22}c_{23}c_{24}}\Big) |c_{1}c_{2}\cdots c_{23}c_{24}> .
  \end{split}
  \end{align}
  $T^1$ lies at the  horizon. $T^4$ , $T^5$ and  $T^6$  are in the region which is on the other side of the horizon as explained in the picture and always traced.  Now to compute the reduce matrix for computing entanglement entropy when we trace out the tensor at  2 and 3 and we  immediately see that the RT surface doesn't cross the horizon. The contribution only comes form cutting the link between 2, 3 and 7. Also when we trace out the tensor at  7 and  we will see that the geodesic wraps the horizon once. The RT surfaces are shown in the figure \ref{fig:btz}.

\subsection{Black hole - a region of high bond dimensions?}

In \cite{HAPPY} and \cite{Qi}, it is observed that a region of high bond dimensions behaves in very similar ways to a black hole, in which the entangling surface for example, would naturally avoid the boundary of this region of high bond dimensions, mimicking the behaviour of the RT surface near a black hole horizon.
In the current construction, the RT surface clearly never cuts through the horizon separating the two halves of the lattice. It is a curiosity how this construction of the BTZ black hole is related to the intuition of the high bond dimensions. 

On closer inspection, we reckon there is an effective region of high bond dimensions realized by the arrangement of the bulk lattice. The key lies in the relation (\ref{trans}). The exponential maps means what appears to be equally spaced in $\phi$ is highly uneven in $\omega$ coordinates, on which the tensor network is based. The bottom line is that comparing the effective length of the first half of the cycle as $\phi\to \phi+\pi$ to the full length of the cycle $\phi \to \phi + 2\pi$, one sees that
\be
\frac{(\omega_{\phi=2\pi}- \omega_{\phi=0})}{(\omega_{\phi=\pi}- \omega_{\phi=0})} = \frac{x^2-1}{x-1} \sim x \gg 1 \,\,\textrm{for} \,\,{x \gg 1}, \qquad x = \frac{2\pi^2}{\beta}.
\ee
This means that traversing half of the $\phi$ cycle still corresponds to an uncannily tiny region in the $\omega$ coordinates, particularly so in the limit of high temperature. The relative number of physical degree of freedom within the region residing in the first half of the space would in fact constitutes a small fraction of the total number of degree of freedom available. There is thus an effective region of high bond dimensions outside of the region, which it has no way of entangling with and the RT surface would thus look very far away from the horizon, which mimics an avoidance of the horizon. We interpret this as a geometrical realization of a region of high bond dimensions. 

\subsection{A comparison with the random tensor construction}

One can readily obtain a general picture of generic wave-functions constructed on these networks with large bond dimension. As discussed in \cite{Qi}, we can obtain a generic picture when bond dimensions are high, in which case the typical behaviour can be deduced by averaging the tensors weighted by Haar measure. 

Let us consider in particular the case where we are computing the Renyi entropy at $n=2$. It is discussed in detail in \cite{Qi} that the  computation of the Renyi entropy can be mapped to that of computing the partition function of an Ising model.

In particular, since upon average of the random tensor located at each node gives
\be
\overline{|V_x\rangle \langle V_x| \otimes |V_x\rangle \langle V_x| } = \frac{I_x + \mathcal{F}_x}{D_x^2 + D_x},
\ee
where $I_x$ corresponds to the identity , where $\mathcal{F}_x$ corresponds to swapping the two copies of the tensors at $V_x$, and $D_x$ the dimension
of the Hilbert space at vertex $x$, which is equal to the product of dimensions of all the bonds that meet at $x$.

The Renyi entropy of some boundary region $A$ is then given by
\be
S_2 \approx  - \log \frac{\bar{Z_1}}{\bar{Z_0}}, \qquad \bar{Z_1} = \sum_{s_x} \exp(- A[\{s_x\}]), 
\ee
with
\be
A = - \sum_{xy} \frac{1}{2} \log D_{xy}(s_x s_y - 1) - \sum_{x\in \partial } \frac{1}{2} \log D_{x\partial} (h_x s_s - 1).
\ee
Here  $s_x \in \{1,-1\}$ are the effective ``spin" degree of freedom which denotes whether the bond at vertex $x$ is connected to the next replica, or with itself.
$h_x$ dictates whether a bond dangling at the boundary should be connected to the same replica for spins outside of region $A$ (in which $h_x=1$) or the other replica for boundary spins inside region $A$, in which $h_x=-1$.  (Note that in our case we have chosen a wave-function that is a pure state in the bulk and thus there is no further contribution from a bulk density matrix as in \cite{Qi}).

We are tracing out everything across the horizon by default to simulate a thermal mixed state. 
In this case therefore, the saddle point contribution in the large $D_x$ limit is such that there are domain walls separating regions with opposite spins that are adjacent to region A at the boundary and its complement, whose effective boundary spin degree of freedom are subjected to opposite directions of the effective magnetic field $h_x$.

\begin{figure}
\centerline{
 \includegraphics[height=5.5cm]{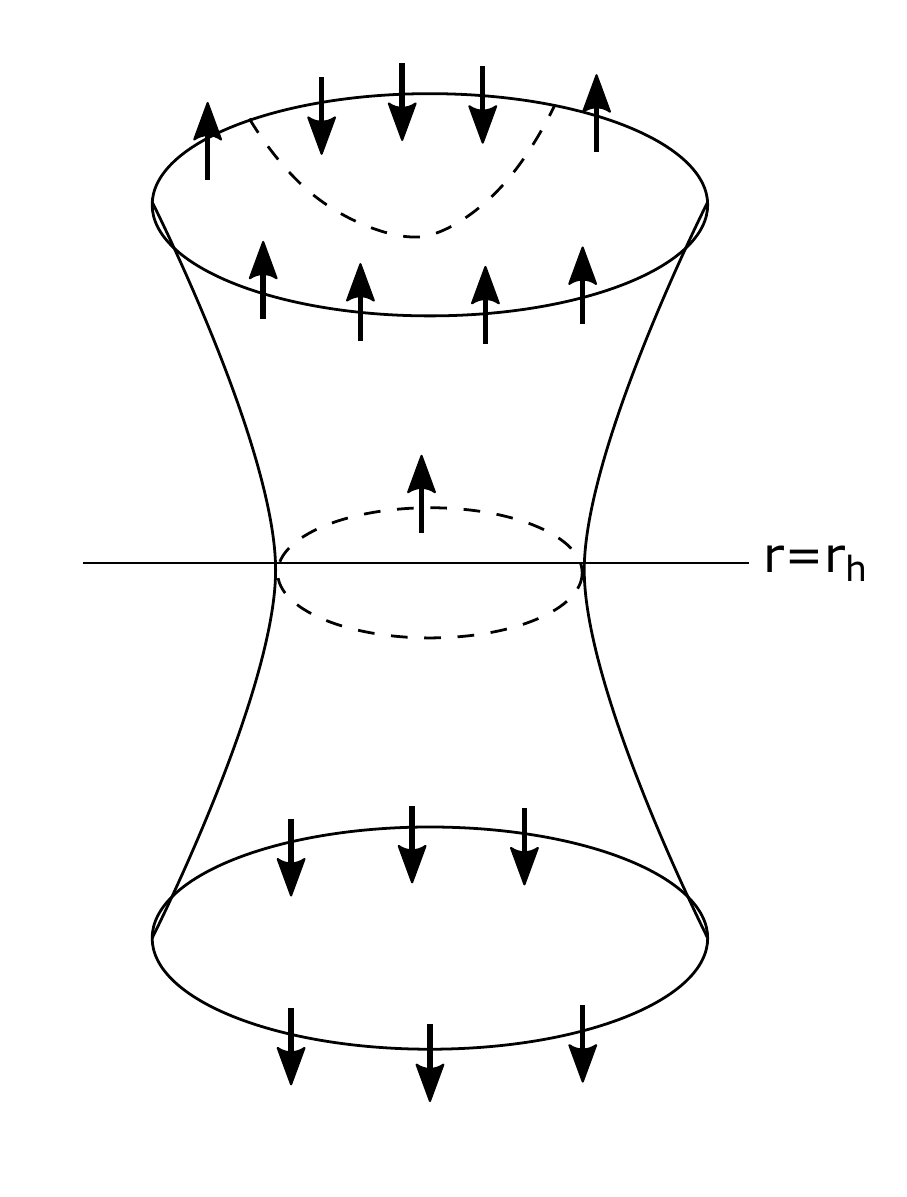}
\hskip 2.cm \includegraphics[height=5.3cm]{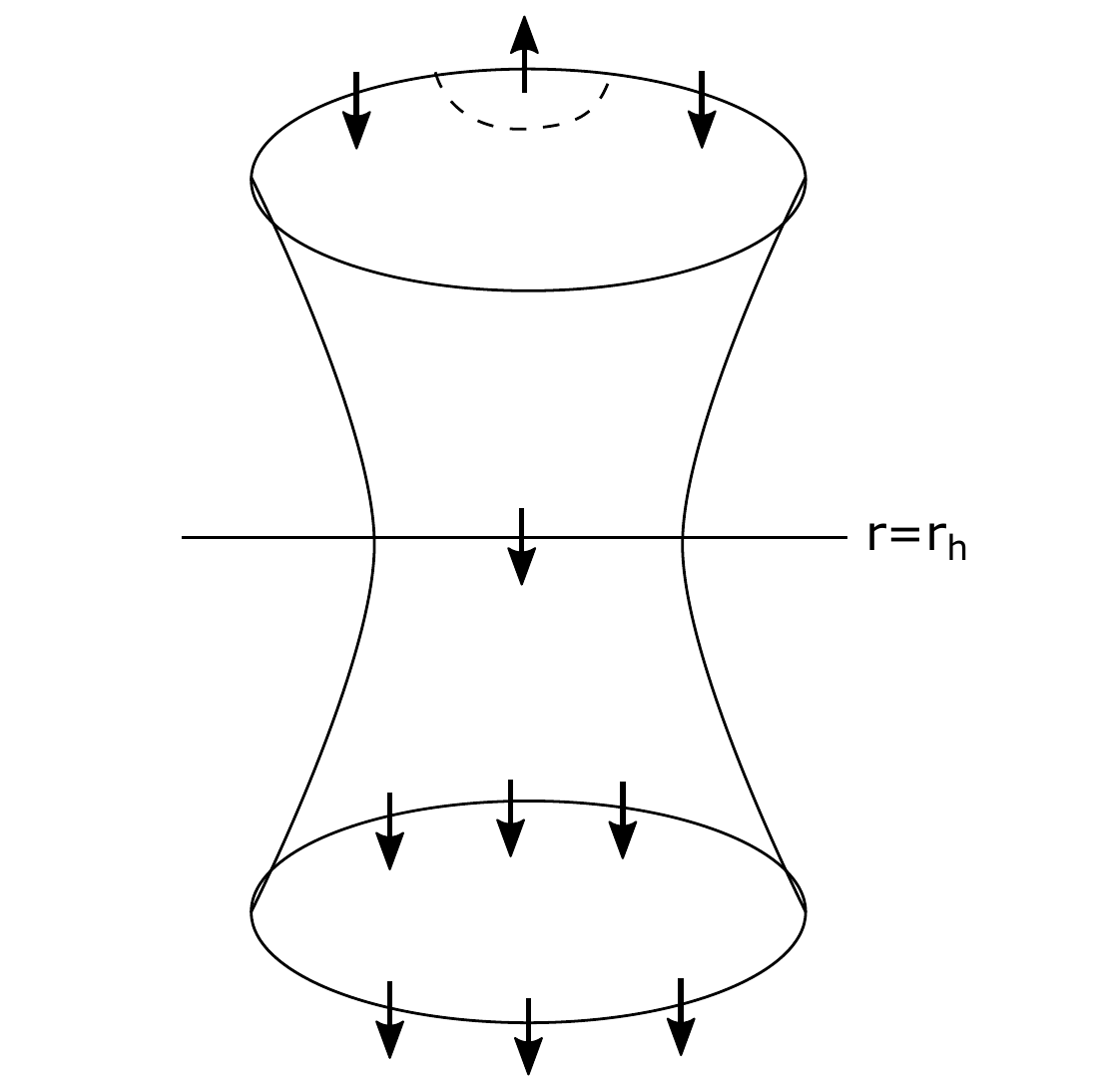}~~~~~~
}

\caption{The left panel illustrates the spin-configuration corresponding to the Renyi entropy of a large region, such that the domain wall covers the ``horizon" as well as its complement. The right panel is the corresponding configuration of the Renyi entropy of a small region, with only a single domain wall. 
}
\label{fig:2spin}
\end{figure}

In this picture, it is then clear that for sufficiently small region $A$, the domain wall again encloses the small boundary region A.  This is illustrated in figure \ref{fig:2spin}. This should be contrasted with the results following from perfect tensors, where we recover also two RT surfaces, one that wraps the horizon and another that does not, depending on the size of the region A.  

\subsection{Computing the thermal spectrum}
In this section we investigate the thermal spectrum. We will use the wavefunction as mentioned in (\ref{wave1}). It can be easily seen the tensor $T^1$ that connects the nodes between the two sides of the horizon  plays the most important role. For simplicity we consider a wavefunction only made up of $T^1$ as follows,
\be
|\psi>=T^1_{\alpha\beta \gamma \gamma \delta\sigma}|\alpha\beta\delta\sigma>.
\ee
We now construct the reduce density matrix using this wavefunction and trace out the region inside the horizon. Then we compute the eigenvalues of these reduce density matrix ($\rho_{\textrm{thermal}}$), that in turn gives the thermal spectrum. 
\be
\rho_{\textrm{thermal}}= \Tr_{\delta\sigma\delta'\sigma'}\Big(T^1_{\alpha\beta \gamma \gamma \delta\sigma}|\alpha\beta \delta\sigma><\sigma'\delta'\beta'\alpha'|T^1_{\alpha'\beta'\gamma'\gamma'\sigma'\delta'}\Big).
\ee
For six index perfect tensor, all the eigenvalues of this density matrix are equal, in fact they are maximally entangled. So the thermal spectra we get is flat like the pure AdS case. Next we consider a bit more non trivial example where there are two nodes on the horizon. The wavefunction is,
\be  \label{wave2}
|\psi>=T^1_{a c d b e f}T_{b c_1d_1 a e_1 f_1}|cdefc_1d_1e_1f_1>.
\ee
The corresponding reduced density matrix is,
\be
\rho_{\textrm{thermal}}= \Tr_{efe_1f_1e_2f_2e_3f_3}\Big(T^1_{a c d b e f}T_{b c_1d_1 a e_1 f_1}|cdefc_1d_1e_1f_1> <c_2d_2e_2f_2c_3d_3e_3f_3| T^1_{a c_2 d_2 b e_2 f_2}T_{b c_3d_3 a e_3 f_3}\Big).
\ee
Now the tensor contractions in (\ref{wave2}) is not unique. We observe that there are two cases. For some specific contractions we again get maximal entanglement  between the four bits. But there exist a particular configuration for which we get  maximal entanglement but only two bits contribute. We then checked for higher number of nodes on the horizon and we reach the same conclusions. There exists always some configurations when all the bits contribute and they became maximally entangled and there exist some configurations when not all the bits contribute but yet the contributing bits are maximally entangled with each other.   Also as we increase the number of such nodes at the horizon it becomes more and more difficult to find these configurations. Almost all the cases all the bits participate and they are maximally entangled. 
\subsubsection{An extra correction in the Renyi entropy} 

The calculation can be inspected in closer detail at arbitrary Renyi index $n$, which is given by  \cite{Qi}\begin{equation}\label{RenyiEntropyApprox}
  S_{n}(A)=\frac{1}{1-n}\mathrm{log}\frac{Z_1^{(n)}}{Z_0^{(n)}}\simeq \frac{1}{1-n}\mathrm{log}\frac{\overline{Z_1^{(n)}}}{\overline{Z_0^{(n)}}},
\end{equation}
where the expression is again applicable only in the large $D$ limit.  When the density matrix corresponding to $n$-replica is averaged over the Haar measure at each vertex, it gives
\begin{equation}\label{RandomTensorAverage}
  \overline{|V_x\rangle\langle V_x|^{\otimes n}}=\frac{1}{C_{n,x}}\sum_{g_x\in Sym_n}g_x
\end{equation}
with $g_x$ the element of the permutation group of order $n$, and $C_{n,x}$ the normalization constant. Then
\begin{equation}\label{z1bar}
\begin{aligned}
  \overline{Z_1^n}&=tr[\rho_P^{\otimes n} \mathcal{C}^n_A \prod_x \overline{|V_x\rangle\langle V_x|^{\otimes n}}]\\
  &=\sum_{\{g_x\}}\frac{1}{C_{n,x}}\mathrm{tr}[\rho_P^{\otimes n}\mathcal{C}^n_A \bigotimes_x g_x]\\
  &=\sum_{\{g_x\}}e^{-\mathcal{A}^{(n)}[\{g_x\}]}
\end{aligned}
\end{equation}
 and
\begin{equation}\label{A=}
  \mathcal{A}^{(n)}[\{g_x\}]=-\sum_{\langle xy \rangle}\mathrm{log}D_{xy}(\chi(g_x^{-1}g_y)-n)-\sum_{x\in\partial}\mathrm{log}D_{x\partial}(g_x^{-1}h_x)+\sum_x\mathrm{log}C_{n,x}
\end{equation}
where $\chi(g_x^{-1}g_y)$ is the number of permutation cycles of the element $g_x^{-1}g_y$, and
\begin{equation*}
  h_x=\begin{cases}
        \, \mathcal{C}_x^n      &   x\in A\\
        \,  \mathcal{I}_x                 &   x\in\overline{A}\\
   \end{cases}
\end{equation*}
where $\mathcal{C}_x^n$ is the cyclic permutation of all the replicas at vertex $x$. 
The $\rho_P$ is the density matrix of bulk states. In the case that has a direct analogue with the tensor network, it is simply given by the expression,  $\rho_P=\otimes\prod_{\langle xy \rangle}|xy\rangle \langle xy|$ where $|xy\rangle$ is an auxiliary state introduced in the ``bulk" corresponding to a maximally entangled state living on the bond that connects the vertices $x$ and $y$. It is maximally entangled to recover simple tensor contraction across a link. In the form of (\ref{z1bar}) and (\ref{A=}), \cite{Qi} suggested that calculating the R\'{e}nyi entropy is the same as calculating the partition function of the spin configuration on the same lattice where spin state was one of the elements of the permutation group, $h_x$ the boundary pinning field, and bond dimension log$D$ the inverse of the temperature. So in the large $D$ limit, the leading contribution of the R\'{e}nyi entropy should come from the ground state configuration of the corresponding spin model with boundary condition \{$h_x$\}. The energy of two adjacent spin was   \[-\mathrm{log}D_{xy}(\chi(g_x^{-1}g_y)-n)\]
So it is a ferromagnetic model, and then it was concluded in \cite{Qi} that the ground state is given by a configuration in which all the spins are parallel to each other, except at the geodesic, where a domain wall is formed between the $\mathcal{C}^n$ area and $\mathcal{I}$ area. The resulting leading term in Renyi entropy is

\begin{equation}\label{RT}
    \begin{aligned}
        S_n(A)  &   \simeq \frac{1}{1-n}|\gamma_A|\mathrm{log}D^{\chi(\mathcal{C}^n)-n}=\frac{1}{1-n}(1-n)|\gamma_A|\mathrm{log}D\\
        &   =|\gamma_A|\mathrm{log}D
    \end{aligned}
\end{equation}
which is the RT formula. However, there is some subtleties in the calculation, which could contribute to some extra non-vanishing terms even in the large $D$ limit.

\begin{figure}
\begin{center}
\includegraphics[width = 0.4\textwidth]{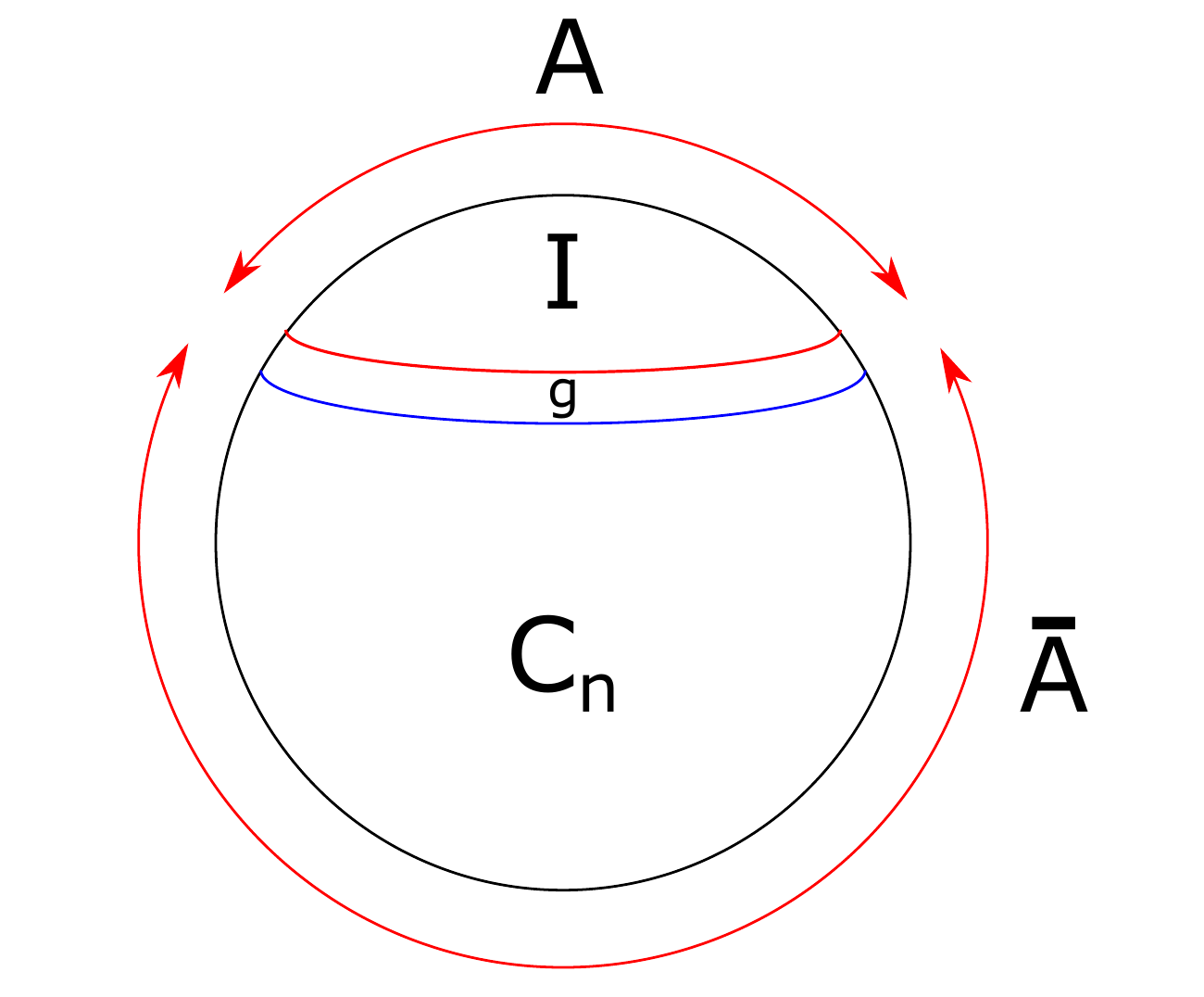}
\end{center}
\caption{A multiple spin domain wall configuration that could have the same ``energy'' as the known contribution with only one domain wall, if the lattice is sufficiently fine-grained.  }
\label{fig:3domain}
\end{figure}

Consider the following: that there are not one but two domain walls, that are very close to each other, both lying very close to the geodesic. The length of each domain wall would almost be $|\gamma_A|$, and in the intermediate area, the ``spins'' are chosen to take value $g$, such that  $g=((12)(3)\dots(n))$, so that $\chi(\mathcal{I}g^{-1})=n-1$ and $\chi(\mathcal{C}^n g^{-1})=2$.  Such configurations are illustrated in figure \ref{fig:3domain}.

The contribution to R\'{e}nyi entropy is
\[
    S'=\frac{1}{1-n}((2-n)+(-1))|\gamma_A|\mathrm{log}D=|\gamma_A|\mathrm{log}D
\]
which is exactly the same as the term in (\ref{RT}). Meanwhile, viewing the cyclic nature of the element $\mathcal{C}$ and $\mathcal{I}$, there are many other possible $g$ that leads to exactly the same action:  
\be g=((1)(23)(4)\dots(n)),((1)(2)(34)...(n))\dots .
\ee 
 Summing all these saddles with equal weight would contribute to a $\frac{\mathrm{log}n}{1-n}$ to the R\'{e}nyi entropy.  It is suppressed compared to the leading term only by a factor of $\log[D]$. 
 
It appears that there are more saddles that could contribute, and the Renyi entropy has more structure less suppressed perhaps than previously thought. 

\section{Conclusion}\label{sec:conclude}

In this paper, we have attempted a remedy of some obvious issues that plague the perfect-tensor-network proposal that attempts to re-enact the AdS/CFT correspondence. These obvious issues include a necessarily flat entanglement spectrum, and the absence of connected correlation functions between local operators. We make a naive proposal that the tensors involved in the tensor network are perturbed away from exact perfection. We demonstrate that such perturbations naturally lead to emergence of some analogues of Witten diagrams,
when we consider 2 pt and 3 pt correlation functions of local operators. We note that these perturbations would also make contributions to the Renyi entropies. It appears that such perturbations naturally lead to non-analytic contribution behaving like $\epsilon\log\epsilon$, if the perturbations away from perfection is orthogonal to the subspace of the boundary Hilbert space the background perfect tensors projects to. To quadratic order they would give rise to non-trivial Renyi entropy, although they are insufficient to fix the issue with the flat spectrum.   We explored some other fix of the entanglement spectrum, which appears to work even though they might be rather contrived. 

We have also constructed the BTZ black hole under the tensor network framework, by orbifolding the discrete lattice characterized by some Coxeter group, analogous to how the BTZ black hole is constructed in the continuous limit. We recover a tensor network that exhibits some salient features of the BTZ black hole, including a horizon which no entangling surface penetrates, and that the entanglement entropy of a region $A$ is different from that of its complement, given by two entangling surfaces one wrapping the horizon while the other does not. The black hole entropy in this case is precisely an entanglement entropy between degrees of freedom across the black hole horizon.

The tensor network proposal has led to quite a few surprises and excitement. The basic feature underlying all of that is the emergence of the notion of a local bulk, which manifests itself as a product of delta functions occurring at each edge whenever we trace out some parts of the boundary of the tensor. It is, as in any other developments, perhaps raising more questions than answers. One important issue is to fix the entanglement spectrum. More importantly, it is necessary to begin discussing time evolution of these networks, in order to actually make contact with other extremely important aspects of the AdS/CFT correspondence. Some systematics such as the handling of isometries is now at hand, using the Coxeter group, and more generally, via these discussions of the PSL$(2,\mathbb{Q}_p)$ groups \cite{Gubser, Caltech}. It appears that time-evolution based on a triangulation of the $AdS_{d+1}$ space and then introducing tensors where some indices are now taken as unitary maps from one time slice to the next is a natural generalization of the current construction of tensor networks on a given spatial slice. We comment that it is observed in \cite{Qi2} that perfect tensors are naturally describing quantum chaotic evolution. Given that the locality of the network and covariance of space-time begs for the use of perfect tensors in the complete space-time network, a coherent picture of gravity, and perhaps black holes in particular being fast scramblers might emerge for free via the tensor network construction. Such construction is currently underway, and we will report them elsewhere \cite{Prog}.   

\section*{Acknowledgements} 
We thank Zheng-Cheng Gu for telling us about the Coxeter group. We also thank Wei Song for very helpful discussions, and for probing us about the BTZ black hole in the tensor network construction which motivated part of this work. 
AB and  LYH would like to acknowledge support by the Thousand Young Talents Program, and Fudan University. ZSG and SNL would like to acknowledge support by Fudan University. AB and LYH would like to thank Sudipta Sarkar and Bin Chen for discussing various relevant issues. Authors would also like to thank Chris Lau for his tremendous help with the figures. 

\begin{figure}
\begin{center}
\includegraphics[width = 0.45\textwidth]{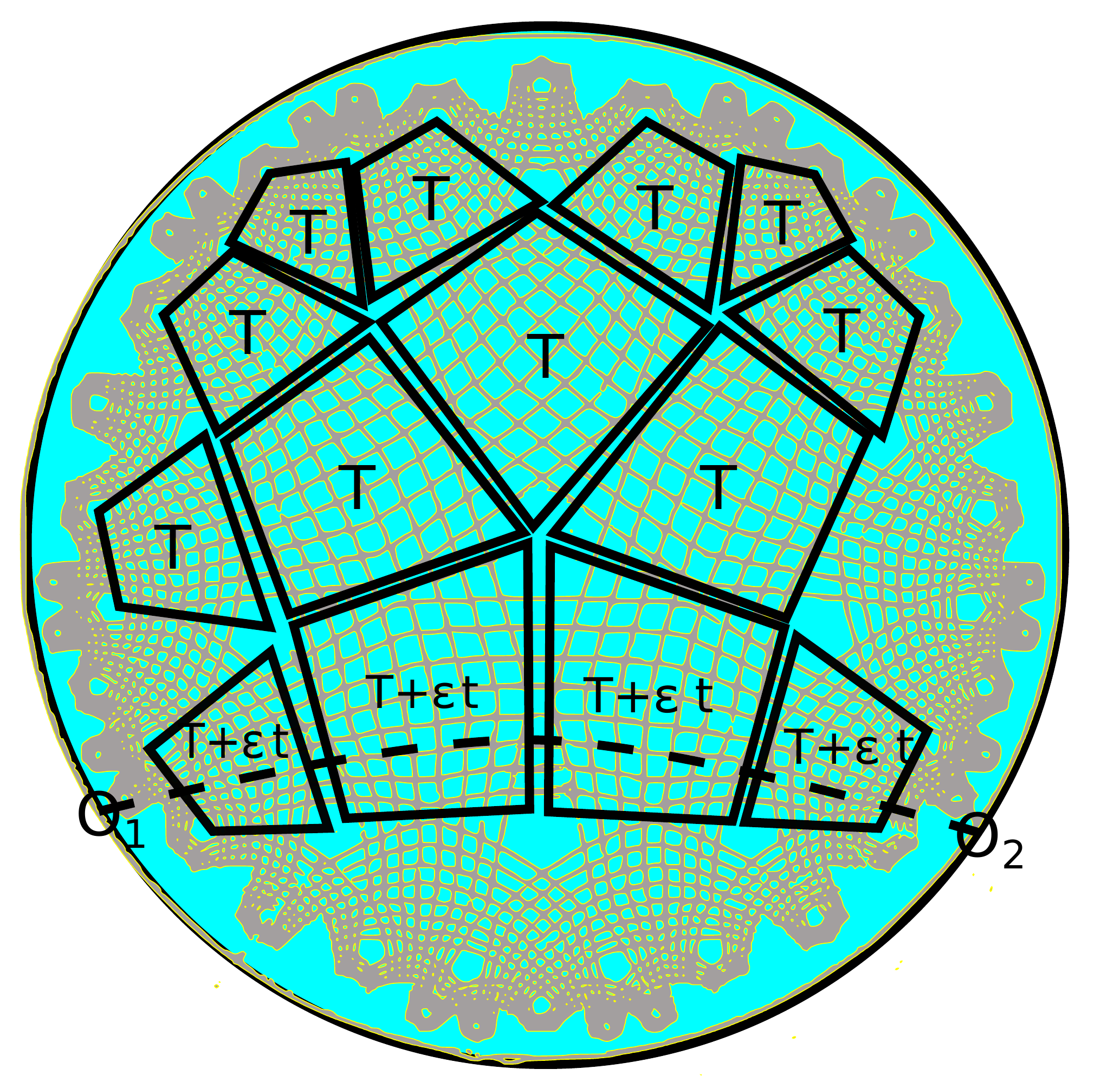}
\end{center}
\caption{The tessellation that appears in \cite{Qi} actually corresponds to a [4,5] lattice. Individual 4-leg tensors cannot be treated as a single unit. The leading contribution in connected correlation functions in this case correspond to a collection of these blocks each with one node replaced by $ t$ forming a connected set of blocks.   }
\label{fig:4leg}
\end{figure}  

\appendix
\section{Curvature and bulk locality}\label{flatness}
In this paper, the main focus is the absence of connected correlation function between local operators. It is found that  departure from perfection in isolated nodes in the tensor network lead only to a change of the overall normalization of operators, but do not contribute to connected correlation functions. That this works actually depend heavily also on the negative curvature of the tessellation. For example, the discussion in \cite{Qi} is based on a 4-leg tensor forming a square lattice. The lattice arranges itself into self-repeating units which can be described by the tessellation [4,5]. The self-repeating units are 4-gons, as indicated in figure \ref{fig:4leg}. \footnote{We adopted the figure from \cite{Qi}.}

\begin{figure}
\begin{center}
\includegraphics[width = 0.40\textwidth]{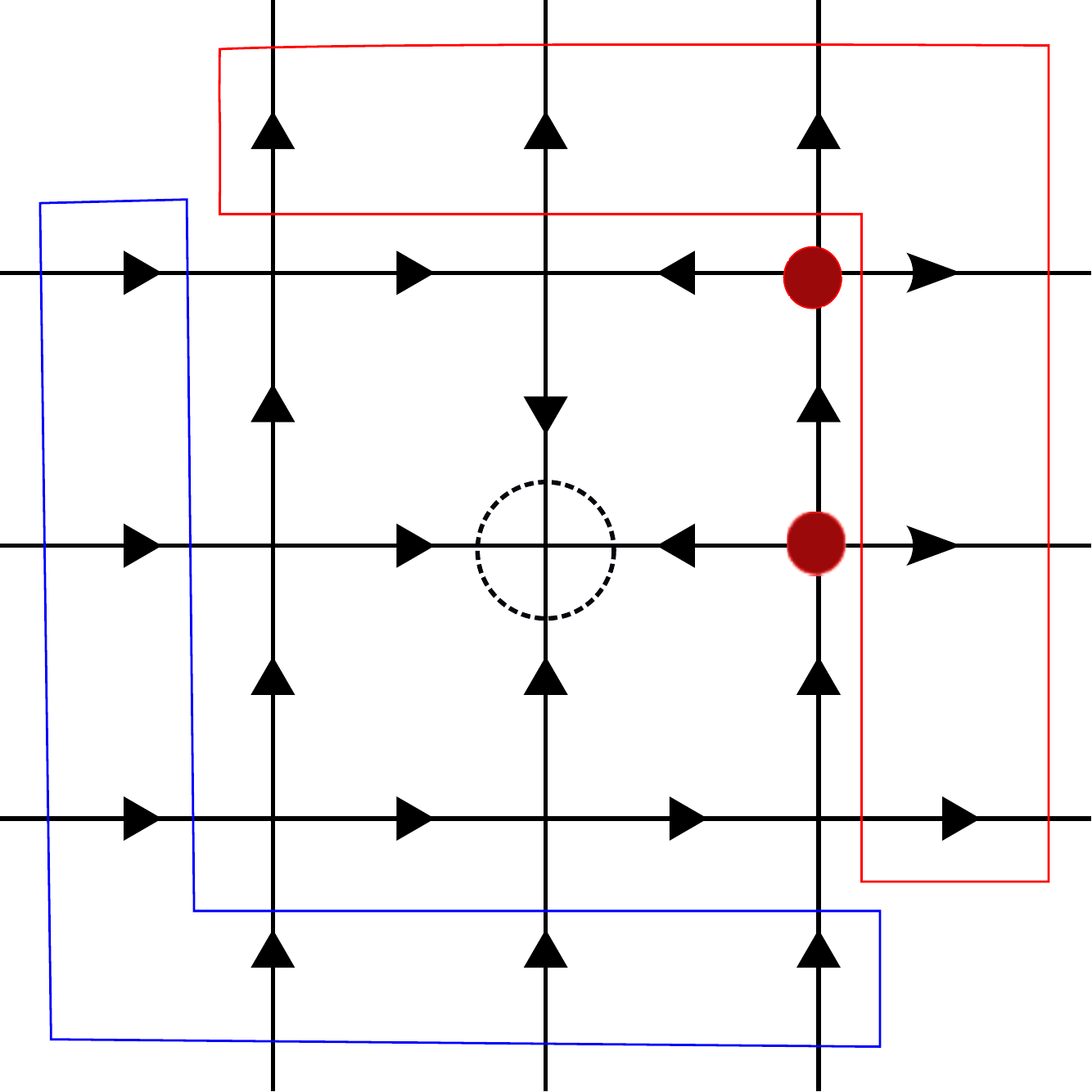}
\end{center}
\caption{Focussing only on one block that looks locally flat, it is clear that one ``hole'' in the middle of the network would destroy the unitary map from the incoming links surrounded by the blue curve to the out-going links surrounded by a blue curve. The red dots denote nodes where the number of incoming arrows and outgoing arrows are not balanced. }
\label{fig:1block}
\end{figure}

A crucial point here is that while the network is built up from 4-leg tensors, the actual unit that defines the geodesics and RT surfaces would be determined by these blocks of tensors, rather than individual 4-leg tensors.  To visualize that, let us focus on one block, depicted in figure \ref{fig:1block}.

Within this block, we could view the left and bottom set of dangling legs as input legs, and the top and right set of dangling legs as output legs in a unitary map defined by the block of 4-leg tensors. Being on a flat geometry means that on average the number of ingoing legs connecting from a layer closer to the boundary dangling legs to out-going legs connecting to an inner layer roughly balances . Suppose we now introduce a hole somewhere inside this block. It is immediately clear from our flow diagram that it is impossible to construct a conserved flow from the input legs that flows out to the output legs and the hole, without either introducing loops in the flow, or having unbalanced in/out legs in a given node. This means that the unitary map of the block is destroyed by introducing only one hole in this block that could contain many 4-leg unit tensor. This follows from the fact that the square lattice has the structure such that a contraction of the outer layers have less than half of the total number of legs, ingoing and outgoing included,  unlike the case of negatively curved lattices. As a result, if we were computing correlation functions in a flat lattice, our previous argument would break down: namely, that a single node perturbed away from perfection would immediately contribute to connected correlation functions. To make connection with our discussion in the tessellation of a negatively curved space, it is thus necessary to view the block of tensors as a single unit, in which case the averaged number of incoming legs from the boundary still vastly exceeds the out-going legs toward the bulk, and our definition of the transfer matrices and fusion matrices would work equally well then.  

\section{Inspecting a remedy to the flat entanglement spectrum-- weighting the legs} \label{flatness}

In the main text, we have seen that the entanglement spectrum is completely flat, which follows from the nature of the perfect tensors. We have also seen that a small deviation from perfection alone does not lead to a satisfactory spectrum. 

In \cite{Qi} it is suggested that perhaps one could introduce non-maximally entangled auxiliary states along the edges connecting vertices. In our tensor network construction, this is equivalent to introducing weights when tensors are contracted.
i.e.

\be
T_{a_1 \cdots} K_{a_1 a_1'} T_{a_1' \cdots},
\ee
for some matrix $K$ satisfying $\tr( K . K^\dag) = 1$.  In \cite{Qi}, it is observed that the averaged tensors with large bond dimension is not very sensitive to such weights. Things are quite different in a given tensor network with specific perfect tensors sitting at each node. While $T. T^\dag = \delta$, $T.K.K^\dag.T^\dag$ generically do not recover delta functions. As a result, the RT surface would fail to emerge in the presence of these weights, unless they are also randomly chosen. 

However, given the negative curvature of the tensor network, one could ask whether it is still possible to change the entanglement spectrum while preserving the RT surface. One observation is that if the weights are not too densely populated -- weights $K$ are sparsely populating the edges, then it is possible that their presence do not destroy the RT surface. The question is: what is the maximum density of weighted edges such that the RT surface remains intact?

We note immediately that in the 4-leg code, locally the geometry is flat. For the same reasons as discussed in section \ref{flatness}, the addition of weights quickly delocalizes the RT surface (i.e. Schmidt-decomposition) into blocks of the self-repeating units. 

It's more instructive to work with the $2L$-leg tensor in which the geometry is manifestly negatively curved. 

A weight on an internal leg prevents the indices of perfect tensor and its adjoint from being contracted directly. Therefore, a weight works as a ``barrier'' obstructing the flow of contraction that propagate from tensor to tensor into the interior of the tensor network as some parts of the physical degrees of freedom is traced out.  The effect of a weight disappears if both nodes connected by this weighted link have at least half of their legs already contracted, so that the weighted link would itself be contracted with a delta function and sum to 1. Making use of this constraint, we can draw some conclusions as follows.
\subsection{The maximal number of weights when pushing toward the inner most layer }
In this section, we study the maximal number of weights without spoiling the propagation of delta function in the next layer $k$ as one contracts some legs connecting to the $k+1$  layer such that the process eventually hits the center of the network, i.e. layer 1.
Suppose there are two tensors, tensor A at Layer $k$ and tensor B at $k+1$, and that they are connected via an internal leg AB. Consider two cases.
\par Case 1: AB is an internal leg with weight. As explained above, we need delta-functions to remove the effect of the weight. That is to say, among the other legs of A and B, there should be at least $L$ (out of the total number of tensor legs $2L$) that have no weights.
\par Case 2: AB is an internal leg without weight. We need $L$ legs, not including AB, without weight to give a delta-function on leg AB. As for tensor A, since we have already had an index contracted, we only need another $L-1$ legs without weight, in order to propagate the delta functions from layer $k$ to layer $k-1$. 

It is thus necessary that each tensor has less than $L$ weighted legs.


\end{document}